\documentclass[11pt]{article}

\usepackage{amsmath,amssymb}
\usepackage{graphicx}
\usepackage{dsfont}
\usepackage{xcolor}

\title{Imaging geometry- and phase-controlled spectra in a surface-state Andreev cavity}
\author{Adrian Greichgauer\textsuperscript{1}, Yoichi Ando\textsuperscript{1} \& Jens Brede\textsuperscript{1,*}\\[0.45em]
\textsuperscript{1}\small Physics Institute II, \href{https://ror.org/00rcxh774}{University of Cologne}, D-50937 K{\"o}ln, Germany
\textsuperscript{*}\small \href{mailto:brede@ph2.uni-koeln.de}{brede@ph2.uni-koeln.de}}
\date{}

\newif\ifnaturestyle
\naturestyletrue          
\usepackage[T1]{fontenc}
\usepackage[utf8]{inputenc}
\usepackage{newtxtext,newtxmath}
\ifnaturestyle
  \usepackage[a4paper,margin=1.6cm,columnsep=6mm]{geometry}
  \AtBeginDocument{\twocolumn}

\else
  \usepackage[a4paper,margin=2.5cm]{geometry}
\fi

\usepackage[super,sort&compress,comma]{natbib}

\usepackage{xurl}

\usepackage[
  colorlinks=true,
  linkcolor=blue,
  citecolor=blue,
  urlcolor=blue
]{hyperref}

\usepackage{geometry}
\begin{document}
\maketitle

\begin{abstract}
Andreev cavities provide a setting in which superconducting proximity spectra are shaped by phase-coherent electron--hole motion along extended trajectories. While such Andreev physics is well established in transport, local spectra in two-dimensional cavities remain largely unexplored in real space. Here we use scanning tunnelling spectroscopy to study confined Cu(111) surface states coupled to superconducting Nb(110). The in-plane magnetic-field scale for the collapse of the resolved low-energy spectrum is controlled by the transverse extent available to Andreev trajectories, while the zero-field excitation energy evolves with the characteristic trajectory length. These trends, together with spatial variations within individual islands and the response to vortex phase textures, are captured by a minimal semiclassical phase-accumulation picture. 
Our results identify geometry-defined Andreev trajectories as a design principle for phase-coherent superconducting cavities accessible by local spectroscopy.
\end{abstract}
\par\vspace{0.5em}
\noindent

\subsection*{Introduction}

Superconducting proximity effects provide a versatile route for engineering hybrid quantum states in mesoscopic or low-dimensional conductors and underpin recent efforts in superconducting electronics, including nonreciprocal Josephson devices, Majorana platforms, and multi-terminal superconducting circuits\cite{Baumgartner2022,Ren2019,Pankratova2020}. Related gap-engineering strategies are also used to control quasiparticle dynamics and dissipation in superconducting devices\cite{Riwar2019}. In these settings, superconducting correlations are expressed through phase-coherent Andreev processes whose spectra depend on geometry, phase bias, and the available electron--hole trajectories.

Andreev cavities provide a particularly direct realisation of this principle. In confined coherent systems, phase-coherent Andreev reflection gives rise to subgap states whose energies depend on path length, boundary scattering, and superconducting phase accumulation\cite{dGSJ1963,Andreev1966,McMillan1968,Kulik1969,Stone1996,Beenakker1997,Mandal2026}.

Scanning tunnelling spectroscopy (STS) has provided direct real-space access to superconducting proximity phenomena in several complementary limits. 
Diffusive SNS weak links have revealed phase-controlled minigap suppression and its local spectral evolution in close agreement with quasiclassical Usadel descriptions\cite{Sueur2008,Roditchev2015}, while cleaner metallic overlayers have shown discrete proximitised subgap states consistent with coherent Andreev bound-state physics beyond a simple diffusive description\cite{Trivini2023,Ortuzar2023}. 
Induced superconductivity in noble-metal surface states has recently been established in atomically engineered quantum dots and confined surface-state islands on Nb(110)\cite{Schneider2023,Schneider2024}. 
An open question is how this physics evolves when confinement is extended beyond few-state structures to larger cavities, where dense ensembles of phase-coherent trajectories may determine the local superconducting spectrum relevant to phase-sensitive hybrid devices.

Here, we use scanning tunnelling spectroscopy of Cu(111) islands on superconducting Nb(110) to realise a phase-tunable Andreev cavity in an extended two-dimensional surface state. The Cu(111) surface state is coupled to the underlying Nb via surface--bulk scattering, so that it inherits an effective superconducting gap scale $\Delta_{\mathrm{eff}}$ from the substrate\cite{Schneider2023,Schneider2024}. The measured island geometry defines the available ballistic trajectories, while an applied in-plane magnetic field tunes the superconducting phase gradient across them.

We find that the low-energy spectrum can be intuitively understood from this combination of island geometry and magnetic-field-controlled phase accumulation. Across multiple islands, the zero-field excitation energy decreases with the length of the relevant trajectories, whereas the field scale of the first zero-bias resonance is set by their projected transverse width. Together, measured island geometries, local spectroscopy and vector-magnet phase control make the Andreev cavity directly addressable in real space and through superconducting phase control, revealing low-energy spectral collapses that follow from a minimal semiclassical phase-accumulation picture.

\subsection*{Island geometry and the magnetic-field response}

\begin{figure*}[ht!]
	\centering
	\includegraphics[width=\textwidth]{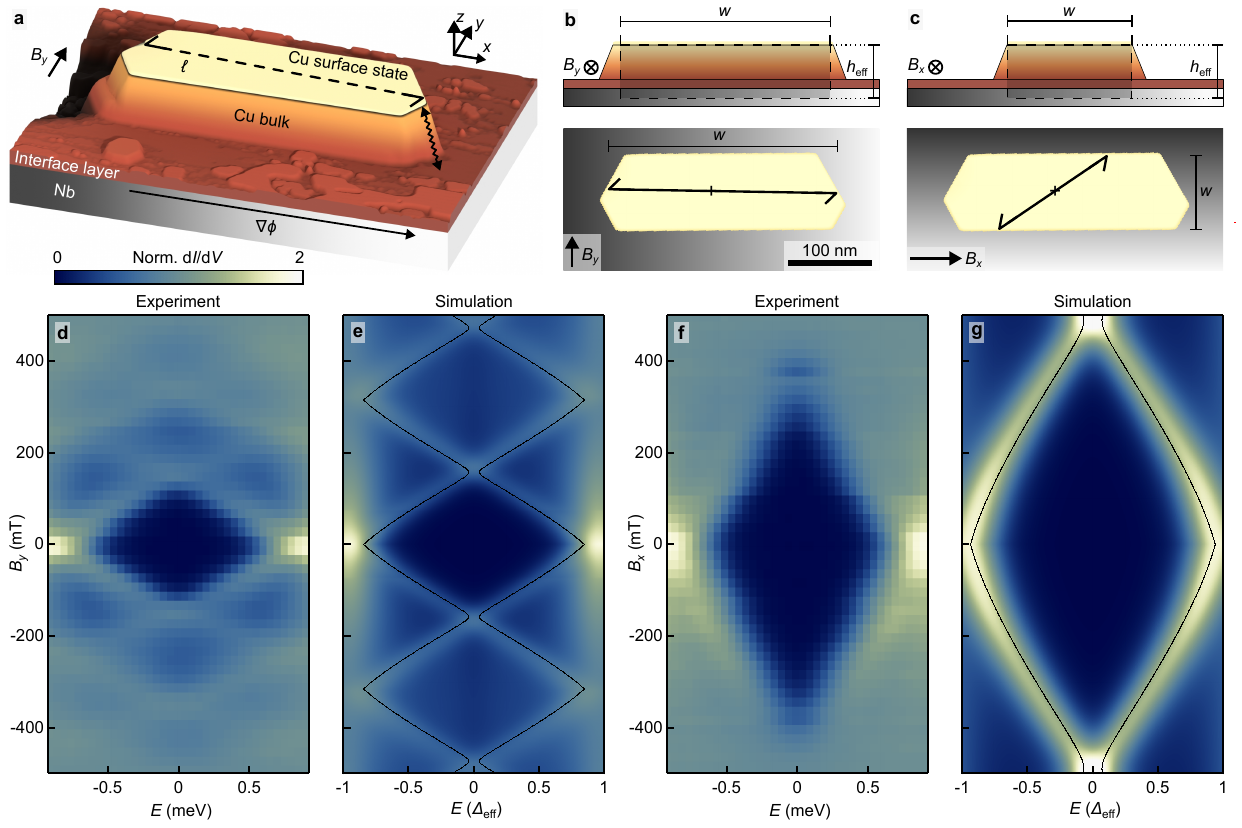}
    \caption{
    \textbf{Island geometry and the magnetic-field evolution of a proximitised Cu(111) surface state.}
    \textbf{a}, Three-dimensional rendering of the STM topography of a Cu(111) island on Nb(110). The wavy arrow marks boundary surface--bulk scattering, giving the confined surface state an effective gap scale \(\Delta_{\mathrm{eff}}\). The dashed black line illustrates a trajectory associated with the length scale \(\ell\). An in-plane field \(B_y\) induces a superconducting phase gradient \(\nabla\phi\) in Nb, indicated schematically by grey shading in \textbf{a--c}.
    \textbf{b,c}, Same island and spectroscopy position, marked by the cross, for fields along \(B_y\) and \(B_x\), respectively. The black line marks an illustrative trajectory associated with the transverse projection \(w\). Together with \(h_{\mathrm{eff}}\), \(w\) defines the effective area \(h_{\mathrm{eff}}w\), which is larger for \(B_y\) than for \(B_x\).
    \textbf{d--g}, Experimental (\textbf{d,f}) and simulated (\textbf{e,g}) differential conductance as a function of energy and magnetic field for \(B_y\) and \(B_x\), respectively. The same island position shows a lower-field collapse of the low-energy spectral gap for \(B_y\) than for \(B_x\), consistent with the larger effective area in \textbf{b}. Black curves in \textbf{e,g} show the calculated bound-state evolution for the highlighted illustrative trajectories. All conductance maps are shown with the same colour scale. The strongest intensities in \textbf{g} exceed the displayed colour scale.
    }
	\label{fig:1}
\end{figure*}

Figure~1 introduces the surface-state Andreev cavity formed by a Cu(111) island on superconducting Nb(110). The island confines the Shockley surface state, a highly mobile two-dimensional surface band residing in a projected bulk gap and directly accessible to the STM tip\cite{Crommie1993}. From spectroscopic measurements, we extract a band bottom \(429\,\mathrm{meV}\) below the Fermi level, an effective mass \(m^\ast\simeq0.45m_e\), and a Fermi velocity \(v_\mathrm{F}\simeq5.8\times10^5\,\mathrm{m\,s^{-1}}\) (Supplementary Fig.~1). In these atomically flat, weakly disordered islands, surface--bulk mixing occurs mainly at the boundary: there, the confined surface state scatters into bulk-like Cu states, hybridises with superconducting Nb, and acquires \(\Delta_{\mathrm{eff}}\approx0.92\,\mathrm{meV}\), consistent with prior STM studies of confined surface states on Nb(110)\cite{Schneider2023,Schneider2024} (Fig.~1a). The boundary therefore both confines the surface-state trajectories and links them to the superconducting phase of Nb. An in-plane magnetic field induces screening currents in Nb and a corresponding superconducting phase gradient \(\nabla\phi\), providing the phase bias sampled by the confined Andreev trajectories. Two geometric quantities are central for each trajectory: how far it travels through the island, and how far it extends transverse to the applied field. The first controls the zero-field energy scale, while the second controls the sensitivity to the field-induced phase variation.

To connect these geometric quantities to the measured island, we extract a trajectory ensemble directly from the STM-defined boundary. From this ensemble, we define trajectory bundles associated with the zero-field spectral scale and with the field scales of the first low-energy spectral collapse. The extraction procedure is shown in Extended Data Fig.~1. The \(\ell\) and \(w\) drawn in Fig.~1 illustrate the geometric scales associated with these bundles, rather than individually selected rays. The same island and spectroscopy position are then measured for two orthogonal in-plane field directions, \(B_x\) and \(B_y\), defined in Fig.~1a. Rotating the field leaves the trajectory lengths unchanged, but changes the projected widths sampled by the trajectories because the field-induced phase gradient rotates with the applied field (Fig.~1b,c). For this elongated island, these projected widths approximately follow the short and long island axes.

This short- versus long-axis difference is directly reflected in the spectra. For \(B_y\), the relevant projected width is long, and low-energy spectral weight appears already at small fields. For \(B_x\), the projected width is much shorter, so the corresponding spectral evolution is shifted to larger fields (Fig.~1d,f). The response is therefore not a geometry-independent suppression of the spectrum, but follows the magnetic phase scale imposed by the island geometry.

\subsection*{A minimal phase-accumulation picture}

This geometry-dependent magnetic-field evolution motivates a simple phase-coherent trajectory description. Because the atomically flat island provides little bulk-like scattering for the surface band, the proximitised two-dimensional surface state can propagate quasi-ballistically across the island. Andreev conversion is then taken to occur at the island boundary, where bulk-derived Cu states link the confined surface trajectory to the superconducting Nb. The amplitude of the proximitising pair potential sets the effective superconducting gap scale, while its phase and the magnetic vector potential enter through the trajectory-dependent phase bias
\begin{equation}
\label{eq:phase_bias}
\delta\varphi =
\int_{\mathbf r_1}^{\mathbf r_2}
\left(
\nabla \phi - \frac{2e}{\hbar}\mathbf A
\right)\cdot \mathrm{d}\mathbf r ,
\end{equation}
where the integrand is the gauge-invariant phase gradient projected along the trajectory; \(\phi\) is the condensate phase, \(\mathbf A\) is the magnetic vector potential, and \(\mathbf r_1\) and \(\mathbf r_2\) are the boundary Andreev conversion points.

In the ideal transparent limit, the corresponding semiclassical Andreev-bound-state (ABS) condition is
\begin{equation}
\label{eq:phase_quantisation}
2\arccos\!\left(\frac{E}{\Delta_{\mathrm{eff}}}\right)
-\frac{2E\ell}{\hbar v_\mathrm{F}}
\mp \delta\varphi
=2\pi \nu,\qquad \nu\in\mathbb Z .
\end{equation}
The three terms are the Andreev reflection phase, the dynamical propagation phase, and the trajectory-dependent phase bias, respectively. At zero field, \(\delta\varphi=0\), solving this condition for the lowest positive energy gives the zero-field spectral scale \(E_0(\ell)\). Thus \(\ell\) controls the dynamical phase accumulated along the electron--hole path and sets the dependence of \(E_0\) on island geometry. In our parameter range, the longest extracted trajectories are comparable to \(\hbar v_\mathrm{F}/\Delta_{\mathrm{eff}}\), placing the islands in an intermediate regime between the short- and long-junction limits: \(E_0\) decreases as longer trajectories become available, but a strict long-junction relation \(E_0\propto1/\ell\) is not expected over the full range.

In an in-plane field, Meissner screening currents in the Nb film generate a slowly varying condensate momentum, equivalently a gauge-invariant phase gradient. This is the Doppler term discussed for proximitised hybrid systems~\cite{Reinthaler2015,Papaj2021}. For the resulting phase bias one obtains
\begin{equation}
\label{eq:inplane_phase}
|\delta\varphi|
\simeq
\frac{2\pi}{\Phi_0}
|B| h_{\mathrm{eff}}w ,
\end{equation}
where \(\Phi_0=h/2e\) and \(h_{\mathrm{eff}}\) is an effective vertical phase lever arm converting the transverse projection \(w\) into the phase-accumulation area \(h_{\mathrm{eff}}w\). Thus \(w\), rather than the full trajectory length, controls the magnetic-field scale. The first zero-bias resonance occurs when this magnetic phase becomes of order \(\pi\), giving
\begin{equation}
\label{eq:B0}
B_0 \simeq \frac{\Phi_0}{2h_{\mathrm{eff}}w},
\end{equation}
and hence the inverse-width scaling \(B_0\propto1/w\).

The single-trajectory estimate sets the geometric field scale. In the full ray-traced island, \(w\) is replaced by the corresponding bundle-derived transverse scale defined in Extended Data Fig.~1. The measured island is not, however, a one-dimensional junction controlled by a single global phase difference. The local spectrum contains an ensemble of trajectories with different lengths \(\ell_n\), transverse projections \(w_n\), transparencies and phase biases. Consequently, the first zero-bias resonance appears experimentally as a gap closing, but it is not followed by a clean reopening as in an ideal single-channel junction. While the largest-\(w_n\) trajectories move away from zero energy, other trajectories with smaller phase bias approach zero and continue to fill the low-energy spectrum.

To account for the imperfect surface-state coupling to Nb, we include a finite interface transparency as a phenomenological parameter, fixed by the experimentally observed zero-bias resonance after experimental broadening. The simulations shown in Fig.~1 and in the following sections therefore extend the minimal phase-quantisation picture to the full measured island geometries, including finite interface transmission and experimental broadening (Extended Data Fig.~2 and Supplementary Information).

The comparison between experiment and simulation in Fig.~1d,e shows that this trajectory-ensemble calculation captures the low-field spectral collapse and geometric field scale, but not the higher-field evolution quantitatively. In particular, the model does not include field-induced weakening of superconductivity in the Nb film, such as gap suppression or phase decoherence. Such effects could shift higher-order resonances to lower absolute fields and further smear the reopening. Separately, substantially lower-field, abrupt or field-asymmetric spectral rearrangements point to changes in the superconducting phase configuration and motivate the later discussion of non-uniform, vortex-like phase textures.

\subsection*{Phase scales across different islands}

\begin{figure*}[ht!]
	\centering
	\includegraphics[width=\textwidth]{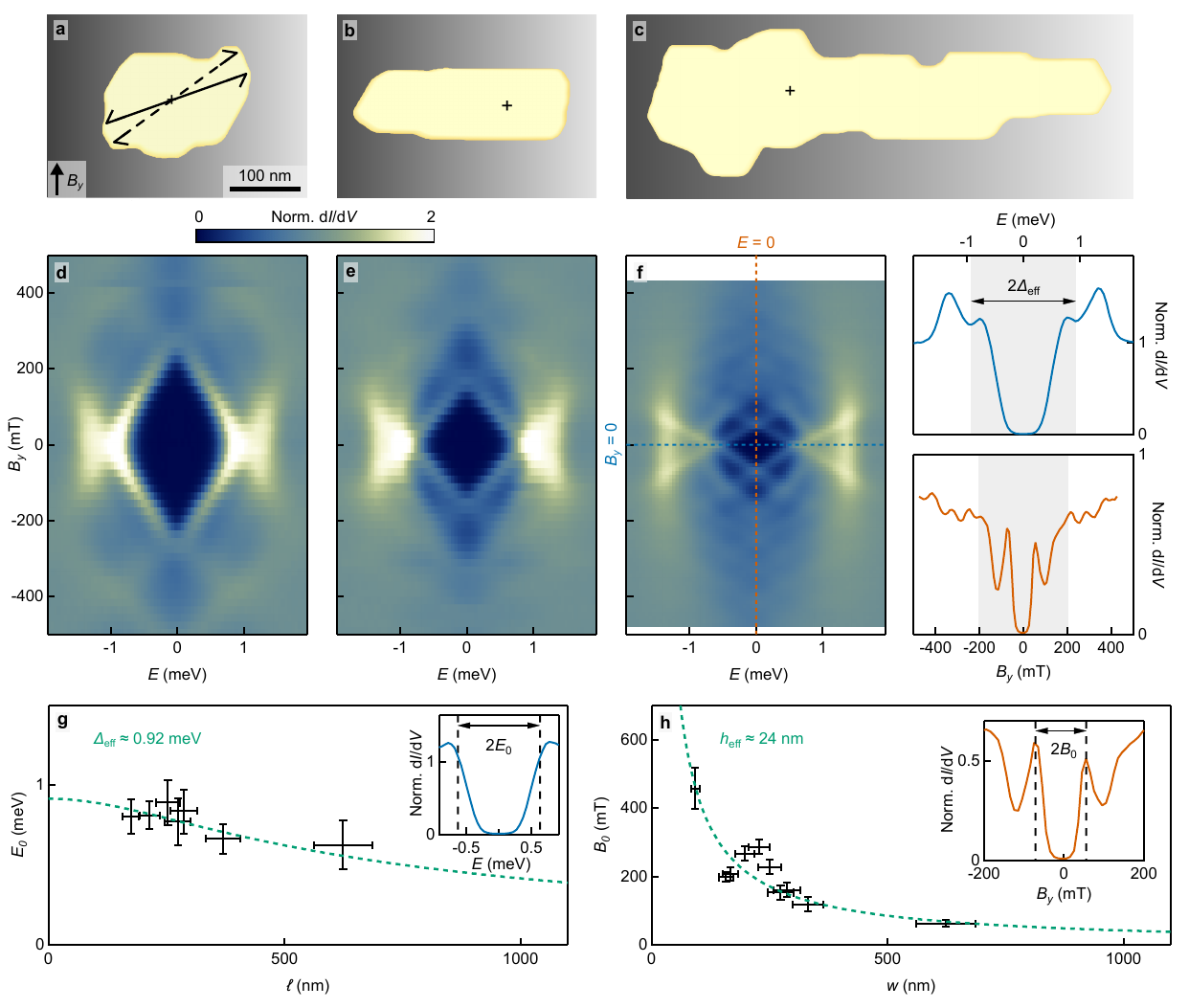}
\caption{
    \textbf{Dynamical and magnetic-field-induced phase scales across different Cu(111) islands.}
    \textbf{a--c}, Representative background-subtracted topographies of Cu(111) islands of different size and shape for magnetic field applied along \(B_y\), shown with a common scale bar. Crosses mark the spectroscopy positions. Grey shading represents the field-induced superconducting phase gradient as in Fig.~1. In \textbf{a}, the solid and dashed arrows indicate trajectories illustrating the transverse projection \(w\) and trajectory length \(\ell\), respectively. The corresponding bundle extraction is defined in Extended Data Fig.~1.
    \textbf{d--f}, Differential-conductance maps as a function of energy and magnetic field for the islands in \textbf{a--c}, showing geometry-dependent evolution of the low-energy spectrum. Line cuts from \textbf{f} are used to extract \(E_0\) and \(B_0\), as illustrated in the insets of \textbf{g,h}.
    \textbf{g}, Lowest excitation energy \(E_0\) as a function of bundle-derived trajectory length \(\ell\). The dashed green line shows the zero-field solution of Eq.~\ref{eq:phase_quantisation} using \(\Delta_{\mathrm{eff}}\approx0.92\,\mathrm{meV}\).
    \textbf{h}, Characteristic field scale \(B_0\) as a function of bundle-derived transverse projection \(w\). The dashed green line shows \(B_0=\Phi_0/(2h_{\mathrm{eff}}w)\), using \(h_{\mathrm{eff}}\approx24\,\mathrm{nm}\).
}
	\label{fig:2}
\end{figure*}

We next ask whether the two geometric scales identified above explain spectra across different Cu(111) island geometries. Figure~2a--c shows islands with different sizes and shapes, all measured for magnetic field applied along \(B_y\). For each spectroscopy position, the measured island boundary defines a ray-traced trajectory ensemble. From this ensemble, we obtain the bundle-derived trajectory length \(\ell\) and transverse projection \(w\), following the procedure illustrated in Extended Data Fig.~1.

The corresponding differential-conductance maps in Fig.~2d--f show geometry-dependent variations of the low-energy spectrum. All three spectra display the phenomenology introduced in Fig.~1: with increasing magnetic field, the low-energy spectral gap evolves towards zero bias and spectral weight is redistributed back to finite energy at larger fields. The field of the first zero-bias resonance varies strongly between islands, already indicating sensitivity to the transverse projection scale.

The characteristic observables are extracted from line cuts through the field-dependent spectra, as illustrated by the insets of Fig.~2g,h. We define \(E_0\) from the lowest-energy spectral feature at zero field and \(B_0\) from the first maximum in the zero-bias conductance. The magnetic-field-axis correction and the extraction of \(E_0\) and \(B_0\) for the measured island set are shown in Supplementary Figs.~2--7, with the resulting island parameters summarized in Supplementary Table~1. Across the measured island set, \(E_0\) decreases with increasing bundle-derived trajectory length \(\ell\) (Fig.~2g), consistent with \(\ell\) controlling the dynamical phase. The dashed line in Fig.~2g shows the corresponding zero-field solution of Eq.~\ref{eq:phase_quantisation}.

The magnetic field scale follows the transverse geometric projection: across the measured island set, \(B_0\) decreases systematically with increasing bundle-derived width \(w\) (Fig.~2h). This is the expected trend for the in-plane-field phase scale, as trajectories with larger transverse projection accumulate the required magnetic phase at lower applied field. The dashed line shows the first-zero-bias condition \(B_0=\Phi_0/(2h_{\mathrm{eff}}w)\), using a single effective height \(h_{\mathrm{eff}}\approx24\,\mathrm{nm}\). This value is consistent with an effective vertical phase lever arm set by the Cu island thickness and screening region in the underlying Nb. Together, Fig.~2g,h show that the dominant spectral and magnetic-field scales follow island geometry through bundle-derived geometric quantities: larger \(\ell\) lowers \(E_0\) through the dynamical phase, while larger \(w\) lowers \(B_0\) through the magnetic phase. Detailed spectral contrast and scatter reflect the full trajectory ensemble, broadening, and experimental resolution discussed in Extended Data Fig.~2.

\subsection*{Position dependence of the local spectra}

\begin{figure}[ht!]
    \centering
    \includegraphics[width=0.495\textwidth]{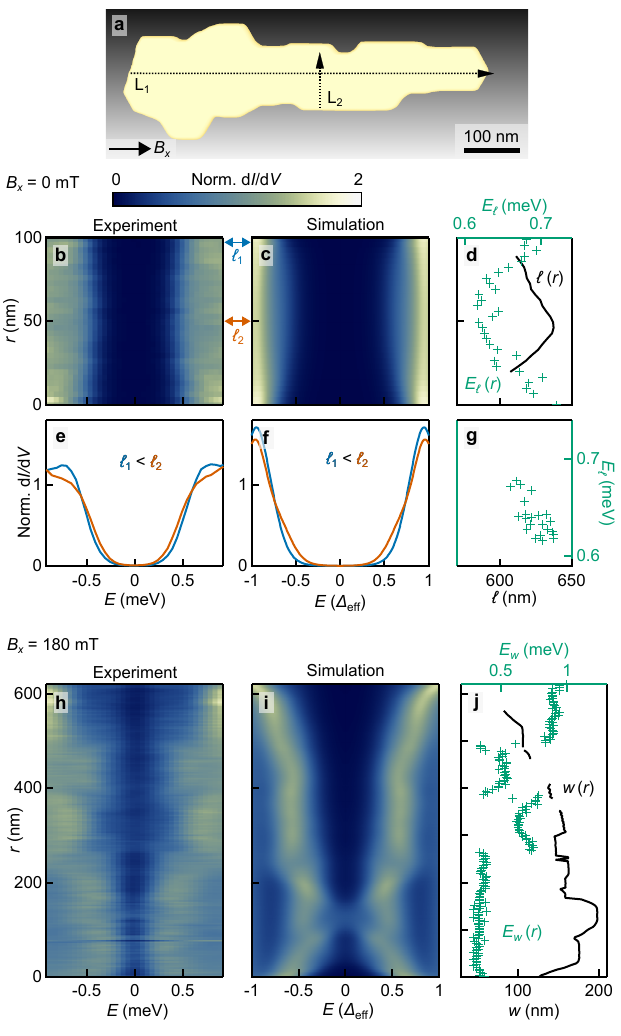}
\caption{
\textbf{Position dependence of the local spectra within a single island.}
\textbf{a}, Background-subtracted topography of the largest Cu(111) island studied, corresponding to Fig.~2c. Grey shading represents the field-induced superconducting phase gradient as in Fig.~1. In \textbf{a}, the solid and dashed arrows illustrate the bundle-derived transverse projection \(w\) and trajectory length \(\ell\), respectively.
\textbf{b,c}, Experimental and simulated differential-conductance spectra along \(L_2\) at zero field.
\textbf{d}, Lowest excitation energy \(E_\ell(r)\) and bundle-derived trajectory length \(\ell(r)\) extracted along \(L_2\).
\textbf{e,f}, Example experimental and simulated zero-field spectra from positions with \(\ell_1<\ell_2\).
\textbf{g}, Correlation between \(E_\ell\) and \(\ell\) along \(L_2\).
\textbf{h,i}, Experimental and simulated differential-conductance spectra along \(L_1\) at \(B_x=180\,\mathrm{mT}\).
\textbf{j}, Finite-field low-energy scale \(E_w(r)\) and bundle-derived transverse projection \(w(r)\) extracted along \(L_1\). Gaps in \(\ell(r)\) and \(w(r)\) mark positions where no trajectory bundle satisfies the minimum-population criterion described in the Supplementary Information.
}
    \label{fig:3}
\end{figure}

Figure~3a shows the largest island studied, with two line-spectroscopy paths probing complementary aspects of the local geometry. This provides an internal test of the geometric picture: within a single island, the local trajectory ensemble changes with tip position while the material parameters and global island boundary remain fixed. For each tip position \(r\), the local ray-traced ensemble is recalculated from the measured island boundary, and the bundle-derived scales \(\ell(r)\) and \(w(r)\) are extracted using the criterion in Extended Data Fig.~1. At zero field, spectra measured along \(L_2\) show a gradual spatial evolution of the low-energy onset (Fig.~3b). The corresponding trajectory-ensemble simulation captures the same trend (Fig.~3c). Extracting the lowest excitation energy \(E_\ell(r)\) along the line shows that positions with larger \(\ell(r)\) have a lower excitation scale (Fig.~3d--g), consistent with the dynamical phase dependence identified across islands in Fig.~2.

In a finite in-plane field, the local magnetic phase scale is probed along the line \(L_1\). For \(B_x=180\,\mathrm{mT}\), the measured spectra show enhanced low-energy spectral weight in the wider part of the island (Fig.~3h), where larger transverse projections \(w(r)\) are obtained. The corresponding simulation captures the same qualitative spatial trend (Fig.~3i). Comparing the extracted low-energy scale \(E_w(r)\) with \(w(r)\) shows the expected tendency towards lower \(E_w\) for larger \(w(r)\), but with substantially more scatter and abrupt spatial changes than in the zero-field comparison (Fig.~3j).

The abrupt changes in experimental \(E_w(r)\) are not captured by the smooth phase-gradient simulation, indicating sensitivity to additional phase structure. Similar discontinuities appear in other field-dependent spectra, predominantly after the initial low-field spectral collapse. A phenomenological comparison in Extended Data Fig.~3 shows how abrupt, asymmetric rearrangements can arise when the smooth in-plane phase term is supplemented by an idealised vortex-like phase texture. This motivates the vortex experiment discussed below.

\subsection*{Effects of vortex phase textures}

We finally probe a qualitatively different superconducting phase texture by applying an out-of-plane field \(B_z=75\,\mathrm{mT}\), which introduces vortices in the underlying Nb film. Because the STM tip probes the Cu(111) surface state rather than the Nb vortex core directly, the vortex is sensed through Andreev coupling to the spatially winding condensate phase of the substrate (Fig.~4a).

\begin{figure}[ht!]
    \centering
    \includegraphics[width=0.45\textwidth]{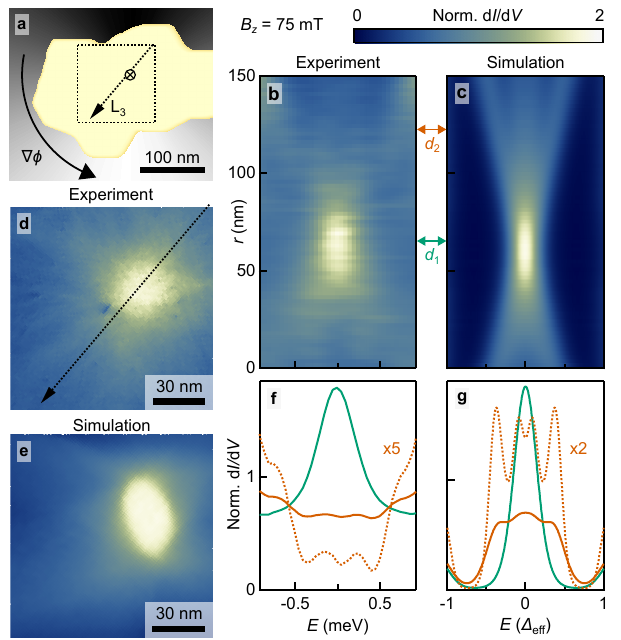}
    \caption{
    \textbf{Local spectral response to a vortex phase texture.}
    \textbf{a}, Zoom into the lower region of the Cu(111) island shown in Fig.~3 under an out-of-plane magnetic field \(B_z=75\,\mathrm{mT}\). The dotted arrow marks the line-spectroscopy path \(L_3\); the dotted square marks the area shown in \textbf{d,e}. Grey shading schematically represents the phase winding around the projected vortex position, marked by the circled cross.
    \textbf{b,c}, Experimental and simulated differential-conductance spectra along \(L_3\).
    \textbf{d,e}, Experimental and simulated zero-bias conductance maps in the region marked in \textbf{a}.
    \textbf{f,g}, Experimental and simulated spectra extracted at the positions indicated in \textbf{b,c}. The labels \(d_1\) and \(d_2\) denote increasing distance from the projected vortex position, with \(d_1<d_2\).
    Dotted curves highlight the four-peak structure: in \textbf{f}, the corresponding experimental spectrum is vertically rescaled and offset. In \textbf{g}, the simulated spectrum is recalculated with reduced effective temperature and vertically rescaled.
    For visual comparison in \textbf{c,e,g}, the simulated intensity is rescaled to the experimental peak amplitude.
    }
    \label{fig:4}
\end{figure}

This distinguishes the present measurement from conventional vortex-core spectroscopy. In a bulk superconductor, the characteristic X-shaped spatial evolution of the local density of states reflects quasiparticles in the core region, where the order-parameter amplitude is suppressed and the gap recovers over the coherence length. Here, the relevant Cu(111)--Nb Andreev coupling occurs away from the strongly gap-suppressed core. We therefore do not interpret the surface state spectra as a direct image of vortex-core gap suppression. Instead, the vortex enters primarily through the phase winding sampled by extended surface-state trajectories.

To understand this at the simplest level, we use an idealised free-vortex phase field centred at the projected phase singularity. In this model, the condensate phase winds by \(2\pi\) around that point, and the resulting phase difference is evaluated for the ray-traced trajectory ensemble using the measured island geometry. This is a simplified limit of the full gauge-invariant phase bias: it retains the phase difference generated by the multivalued condensate phase, but neglects the vector-potential contribution from screening currents.

The resulting spectra (Fig.~4c,e,g) qualitatively capture the main experimental features (Fig.~4b,d,f), including the outer X-shaped evolution and the additional low-energy structure. In the phase-accumulation picture, trajectories passing near the projected vortex centre sample phase differences approaching \(\pi\), shifting Andreev levels towards zero energy. Farther from the vortex, the sampled phase difference decreases and the corresponding features move back to higher energy. The idealised free-vortex simulation places the inner features at lower energy than observed experimentally, so that they merge into a single broadened peak after applying the experimental resolution. The dotted curve in Fig.~4g shows the same simulated spectrum with reduced broadening, revealing the underlying split branches.

This energy offset is consistent with the simplifications of the free-vortex model. By neglecting the vector-potential part of the gauge-invariant phase bias, the model overestimates the phase difference sampled by the trajectories. Including screening currents would reduce the effective phase bias and shift the corresponding Andreev features outward in energy. Thus, Fig.~4 supports that the proximitised Cu(111) surface state responds to vortex phase textures through extended trajectory-dependent phase accumulation, rather than through a direct local image of the vortex-core spectrum.

\subsection*{Discussion}

Our measurements establish proximitised Cu(111) islands as two-dimensional surface-state Andreev cavities, in which local bound-state spectra probed by STS reflect phase accumulation along trajectories spanning the island geometry. A minimal phase-accumulation picture captures the field-orientation dependence in a single island, the cross-island scaling of \(E_0\) and \(B_0\), the spatial evolution within one island, and the qualitative response to a vortex phase texture. The resulting picture links local spectroscopy to island-scale geometry: the low-energy response measured at the tip reflects phase accumulation along extended Andreev trajectories set by the measured island boundary.

This trajectory-based view connects naturally to established descriptions of superconducting proximity systems. In diffusive SNS structures, quasiclassical approaches based on Green's functions successfully describe phase-controlled minigaps and local spectral evolution after averaging over impurity-scattered paths\cite{Belzig1999,Sueur2008,Roditchev2015}. Here, atomically flat Cu(111) islands provide a complementary limit: the confined two-dimensional surface state propagates essentially ballistically, so the low-energy response remains sensitive to coherent phase accumulation along the available trajectory geometry.
Residual scattering, finite linewidths and coupling between trajectories broaden the branch-resolved structure into spectral envelopes, but do not erase the underlying geometric phase scales.

Several comparisons support this coherent-propagation interpretation. First, a one-dimensional diffusive Usadel equation captures the generic suppression of a proximity minigap, but yields a smoother spectral evolution than observed experimentally (Supplementary Fig.~8). This indicates that diffusive averaging alone does not account for the structured low-energy response.

Second, large-bias spectra show that bulk-derived Cu states contribute appreciably to the tunnelling conductance (Supplementary Fig.~9). Applying the same trajectory-scale analysis to a three-dimensional bulk-derived channel shows that these states are also governed by geometric phase scales, but by different ones. The weighted bulk ensemble is dominated by short trajectories set primarily by the island height, rather than by the long in-plane lengths and transverse projections of the confined two-dimensional surface-state trajectories. Together with the larger Nb gap scale, this shifts the bulk-derived response toward the Nb gap edge.
The bulk-derived channel can therefore contribute to the measured conductance without setting the initial low-field collapse. That scale remains controlled by the two-dimensional surface-state ensemble (Supplementary Fig.~10).

Moreover, the vortex experiment in Fig.~4 shows that the trajectory framework also applies when the superconducting phase varies locally. The strongest spectral shifts occur where island-spanning trajectories pass closest to the projected vortex singularity, providing spatial sensitivity to the underlying phase winding. The phenomenological comparison in Extended Data Fig.~\ref{fig:ED_PhaseJump} makes a related point for the abrupt, field-asymmetric changes seen in in-plane field: such behaviour is compatible with local rearrangements of the superconducting phase configuration due to an accidentally created vortex.

This phase sensitivity is relevant beyond the present islands. In-plane fields generate condensate-momentum phase shifts for Andreev trajectories, equivalent to flux through an area \(h_{\mathrm{eff}}w\). This is closely related to the recently reported hybrid nano-SQUID device based on topological-insulator surface states, where an in-plane field tunes the superconducting phase through an effective device area and is used to access a topological regime\cite{Nikodem2025}.

The Cu(111)/Nb(110) platform therefore extends superconducting surface-state spectroscopy to two-dimensional Andreev cavities with atomically defined boundaries and locally tunable geometry. Together with established surface-state patterning by atom manipulation, molecular nanoarchitectures, and self-assembled networks, this approach offers routes to tune trajectory length, scattering, transparency and phase bias, while resolving in real space how these ingredients shape cavity ABS spectra from a simple trajectory-dominated regime toward Andreev quantum-dot and billiard regimes\cite{Altland1996,Zulaica2022}.


\section*{Methods}

\subsection*{Sample preparation}

The a-plane sapphire substrates, \(\mathrm{Al_2O_3}(11\bar{2}0)\), were prepared ex situ by ultrasonic cleaning in acetone and isopropanol, followed by reactive-ion etching in \(\mathrm{O_2}\), \(\mathrm{Ar}\), and \(\mathrm{CF_4}\). After transfer into ultrahigh vacuum with a base pressure below \(5 \times 10^{-10}\,\mathrm{mbar}\), \(100\,\mathrm{nm}\)-thick epitaxial \(\mathrm{Nb}(110)\) films were grown by electron-beam evaporation at \(1\,\mathrm{nm\,min^{-1}}\). The first Nb layers were deposited at a substrate temperature of \(600\)--\(700\,^{\circ}\mathrm{C}\) to promote continuous film formation, followed by deposition at \(1000\)--\(1100\,^{\circ}\mathrm{C}\) and annealing for \(6\)--\(12\,\mathrm{h}\), yielding the oxygen-reconstructed \(\mathrm{Nb}(110)\) surface used as the superconducting template\cite{Razinkin2010}.

Cu was deposited in situ onto this surface by thermal evaporation at \(0.5\,\text{\AA}\,\mathrm{min}^{-1}\). During the first deposition step, the substrate temperature was rapidly increased to approximately \(700\,^{\circ}\mathrm{C}\) and then reduced to \(200\,^{\circ}\mathrm{C}\). After a \(2\,\mathrm{h}\) deposition, the sample was annealed for another \(16\,\mathrm{h}\). A second Cu deposition step was then performed at approximately \(150\,^{\circ}\mathrm{C}\) for \(100\,\mathrm{min}\), after which the sample was transferred in situ to the STM and cooled to \(400\,\mathrm{mK}\).

\subsection*{STM/STS measurements}

All STM experiments were performed in a commercial low-temperature ultrahigh-vacuum STM system (Unisoku USM1300, \(2\text{--}2\text{--}9\,\mathrm{T}\) vector magnet) with a base temperature of \(400\,\mathrm{mK}\). Topographic images were acquired in constant-current mode. Differential-conductance spectra were measured by disabling the feedback loop after stabilising the tip at the chosen setpoint ($V_\mathrm{stab}$, $I_\mathrm{stab}$) and recording \(\mathrm{d}I/\mathrm{d}V\) with a lock-in amplifier at \(311\,\mathrm{Hz}\).
The following parameters were used to acquire the data shown in the main figures:
\\
Fig.~1a--c: \(V{=}900\)\,mV, \(I{=}1.1\)\,nA;
\\
Fig.~2a--c,3a,4a: \(V{=}900\)\,mV, \(I{=}1\)\,nA;
\\
Fig.~1d,f,2d,f,3b,e,h,4b,f: \(V_\mathrm{stab}{=}500\)\,mV, \(I_\mathrm{stab}{=}500\)\,nA, \(V_\mathrm{mod}{=}50\)\,µV;
\\
Fig.~2e: \(V_\mathrm{stab}{=}900\)\,mV, \(I_\mathrm{stab}{=}500\)\,nA,
\(V_\mathrm{mod}{=}50\)\,µV;
\\
Fig.~4d: \(V_\mathrm{stab}{=}500\)\,mV, \(I_\mathrm{stab}{=}500\)\,nA, \(V_\mathrm{mod}{=}100\)\,µV.

PtIr and electrochemically etched W tips were cleaned by Ar sputtering and electron bombardment, and were conditioned on Cu(111) until clean surface-state spectra were obtained\cite{Schmucker2012}.

\subsection*{Trajectory-ensemble simulations}

Trajectory-ensemble simulations were performed using custom Igor Pro code. Island boundaries were extracted from background-subtracted STM topographies and converted into binary masks. For each spectroscopy position, straight island-spanning trajectories were sampled over propagation angle within the measured island geometry. For every trajectory \(n\), the chord length \(\ell_n\), transverse projection \(w_n\) perpendicular to the applied in-plane field, boundary intersection points, and effective interface transparency were recorded. The full simulated local density of states was calculated from the complete trajectory ensemble. The unindexed quantities \(\ell\) and \(w\) used in the scaling plots denote bundle-derived length and transverse-projection scales extracted from the same ensembles.

For each trajectory, bound-state energies were calculated from a finite-transparency ABS condition,
\begin{equation*}
\cos\!\left[
2\arccos\!\left(\frac{E}{\Delta_{\mathrm{eff}}}\right)
-
\frac{2E\ell_n}{\hbar v_\mathrm{F}}
\right]
=
(1-T_n)+T_n\cos\!\left(\delta\varphi_n\right).
\label{eq:methods_finite_transparency}
\end{equation*}
Here \(T_n\) is the effective trajectory transparency and \(\delta\varphi_n\) is the trajectory-dependent gauge-invariant phase bias sampled between the two Andreev-conversion points. For an in-plane magnetic field without additional phase textures, we used
\begin{equation*}
\left|\delta\varphi_n\right|
=
\frac{2\pi}{\Phi_0}
|B|h_{\mathrm{eff}}w_n ,
\end{equation*}
where \(h_{\mathrm{eff}}\) is an effective vertical scale that converts the transverse projection \(w_n\) into the effective area \(h_{\mathrm{eff}}w_n\) for in-plane-field phase accumulation, and \(\Phi_0=h/2e\). Finite normal reflection at the island boundary was parameterised by a BTK-like barrier strength \(Z\), which determines \(T_n\). Unless stated otherwise, simulations used
\begin{equation*}
\begin{aligned}
\Delta_{\mathrm{eff}} &= 0.92\,\mathrm{meV}, &
h_{\mathrm{eff}} &= 24\,\mathrm{nm}, &
E_\mathrm{F} &= 0.429\,\mathrm{eV}, \\
m^\ast &= 0.45m_e, &
Z &= 0.05 .
\end{aligned}
\end{equation*}
These parameters correspond to
\begin{equation*}
v_\mathrm{F}\simeq5.8\times10^5\,\mathrm{m\,s^{-1}},
\qquad
\lambda_\mathrm{F}\simeq2.8\,\mathrm{nm} .
\end{equation*}
A Lorentzian linewidth was applied to the calculated spectra, followed by convolution with the experimental energy resolution set by the effective electronic temperature \(T_{\mathrm{eff}}=0.86\,\mathrm{K}\) and lock-in modulation, as calibrated in Supplementary Fig.~11. Unless stated otherwise in the figure captions, simulated spectra are normalised to the above-gap conductance.

Additional phase textures were included phenomenologically through an extra trajectory-dependent phase contribution, which can be used either alone or together with the smooth in-plane-field phase term. For the free-vortex calculations, this contribution was evaluated as the wrapped phase difference of a multivalued condensate phase between the two trajectory endpoints. This provides a minimal phase-bias model for testing how the trajectory ensemble responds to phenomenological condensate-phase configurations, without requiring a self-consistent solution of the superconducting current distribution.

Further implementation details, including trajectory sampling, transparency weighting, broadening, phase-texture implementation and diffusive Usadel benchmarks, are given in the Supplementary Information.

\subsection*{Data analysis}

STM and STS data were processed using custom Igor Pro routines and the SIDAM package for spectroscopic-imaging analysis\cite{SIDAM}. Magnetic-field values were calibrated for magnet hysteresis as described in the Supplementary Information. Superconducting spectra were normalised by a smooth normal-state background.

\subsection*{Data availability}
The data that support the findings of this study are available from the corresponding author upon reasonable request. Source data will be deposited in a public repository before publication.

\subsection*{Code availability}
The Igor Pro routines used for trajectory extraction, bound-state calculations and LDOS-map generation are available from the corresponding author upon reasonable request. The code is maintained in a GitHub repository and will be made publicly accessible before publication.


\section*{Acknowledgements}
This work was funded by the Deutsche Forschungsgemeinschaft (DFG, German Research Foundation) through CRC 1238-277146847 (subprojects A04, B01 and B06) and through Germany's Excellence Strategy -- Cluster of Excellence Matter and Light for Quantum Computing (ML4Q) EXC 2004/2-390534769. 
The authors also acknowledge support from the DFG Major Instrumentation Programme under project No. 544410649.
We are grateful to Jakob Schluck, Oliver Breunig and Christian Dickel for many insightful discussions throughout the development of this work, and for careful feedback that improved the presentation and interpretation of the results.

\section*{Author contributions}
J.B. and Y.A. conceived and supervised the project. J.B. and A.G. performed STM/STS experiments, analysed the experimental data, and developed the phase-accumulation model. J.B. implemented the trajectory-ensemble calculations and numerical simulations. J.B. wrote the original draft with substantial contributions from A.G. and Y.A. All authors contributed to the scientific interpretation and framing of the results and to the editing of the manuscript.

\subsection*{AI use disclosure}
OpenAI ChatGPT was used to assist with code development and language editing. All code, analyses and text were checked by the authors, who take full responsibility for the final manuscript.

\section*{Competing interests}
The authors declare no competing interests.

\section*{Additional information}
Supplementary information is available for this paper.\\
Correspondence and requests for materials should be addressed to J.B.


\providecommand{\noopsort}[1]{}\providecommand{\singleletter}[1]{#1}%


\clearpage
\setcounter{figure}{0}
\renewcommand{\figurename}{Extended Data Fig.}
\renewcommand{\thefigure}{\arabic{figure}}

\begin{figure*}[t]
    \centering
    \includegraphics[width=\textwidth]{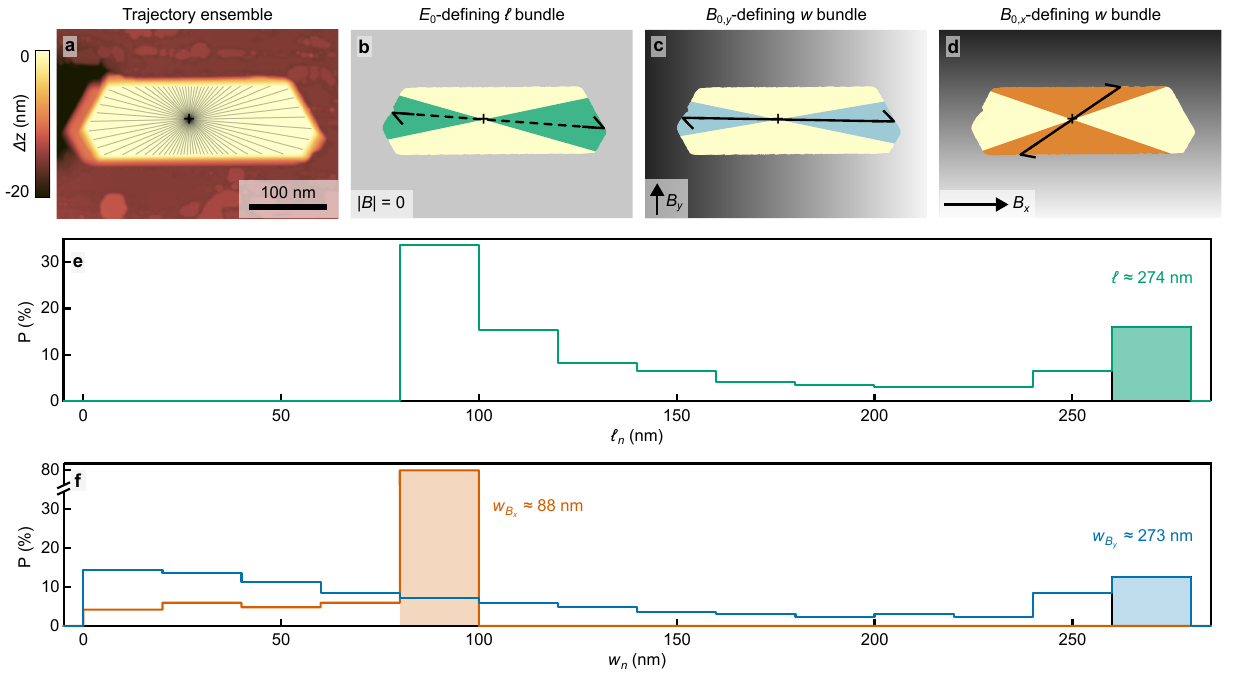}
\caption{\textbf{Extraction of trajectory bundles from the ray ensemble.}
\textbf{a}, Ray-traced trajectory ensemble for the spectroscopy position marked by the black cross, shown on the same STM topography as in Fig.~1. For clarity, only every fifth ray is displayed.
\textbf{b}, Trajectory bundle defining the zero-field spectral scale \(E_0\), obtained by binning the trajectory lengths \(\ell_n\).
\textbf{c,d}, Trajectory bundles defining the field scales \(B_{0,y}\) and \(B_{0,x}\), respectively. For each trajectory, the magnetic phase is set by the projected width \(w_n\) along the field-induced phase-gradient direction, which rotates with the applied in-plane field. Therefore \(B_y\) and \(B_x\) define different \(w_n\) bundles from the same ray ensemble.
The black lines in \textbf{b--d} indicate the bundle-derived geometric scales whose values are annotated in \textbf{e,f}. Grey shading indicates the zero-field reference in \textbf{b} and the field-induced phase gradients in \textbf{c,d}.
\textbf{e}, Histogram of trajectory lengths \(\ell_n\). The shaded bin marks the \(E_0\)-defining \(\ell\) bundle; the annotation gives the corresponding bundle-derived \(\ell\) value used in Fig.~2.
\textbf{f}, Histograms of projected widths \(w_n\) for fields along \(B_x\) and \(B_y\). Shaded bins mark the corresponding \(B_{0,x}\)- and \(B_{0,y}\)-defining \(w\) bundles; annotations give the corresponding bundle-derived \(w\) values used in Fig.~2.
Histograms are normalized to the total number of rays. \(P\) denotes the percentage of trajectories in each bin.
Trajectories are grouped into \(20\,\mathrm{nm}\) bins in either \(\ell_n\) or \(w_n\), motivated by a clean-limit estimate using \(L_\phi\simeq10\,\mu\mathrm{m}\), corresponding to an intrinsic geometric broadening of order \(20\,\mathrm{nm}\) for the relevant cavity trajectories. Only bins containing a minimum number of trajectories are retained as trajectory bundles. The exact criterion is given in the Supplementary Information.
}
    \label{fig:ED_ray_tracing}
\end{figure*}

\begin{figure*}[t]
    \centering
    \includegraphics[width=\textwidth]{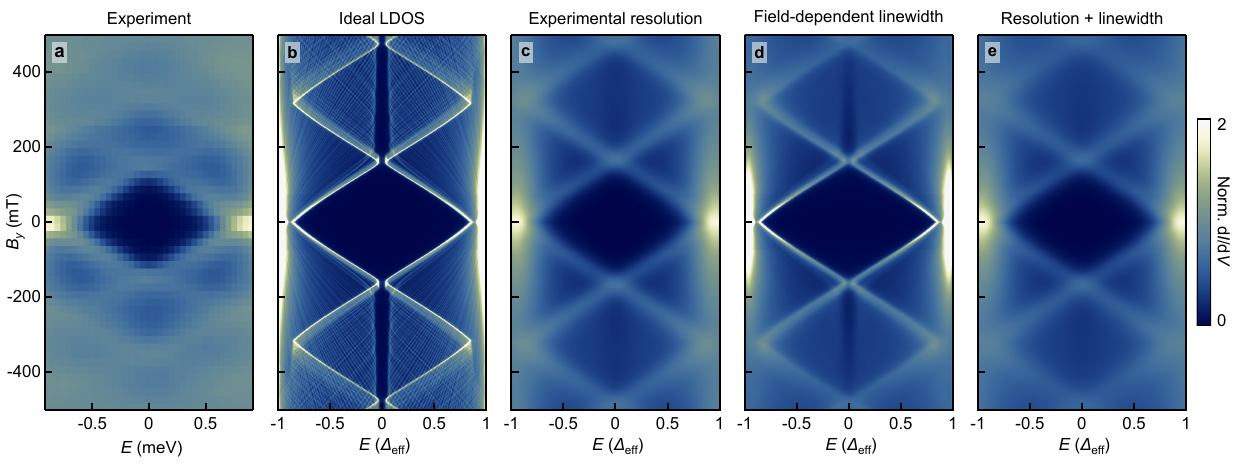}
\caption{
\textbf{Effect of broadening on the trajectory-ensemble LDOS.}
\textbf{a} Experimental spectrum reproduced from Fig.~1 for comparison.
\textbf{b} Ideal trajectory-ensemble LDOS calculated from the ray-traced island geometry. Individual branches should not be interpreted as uncoupled exact eigenstates. Hybridisation and spectral-weight redistribution are expected in a full quantum treatment.
\textbf{c} Convolution with the experimental energy resolution suppresses much of the branch-level contrast while preserving the low-energy spectral envelope and characteristic field scale.
\textbf{d} Adding a phenomenological field-dependent linewidth, \(\Gamma(B)=\Gamma_0+\gamma B^2\), further reduces sharp trajectory-resolved contrast, particularly at larger fields.
\textbf{e} Combining experimental resolution and field-dependent linewidth yields a spectrum closer in appearance to the measured data, while leaving the geometric envelope and field scale largely unchanged. The main-text simulations use the resolution-convolved trajectory-ensemble LDOS in \textbf{c} as the minimal comparison without additional field-dependent broadening.
}
    \label{fig:ED_broadening}
\end{figure*}

\begin{figure*}[t]
    \centering
    \includegraphics[width=\textwidth]{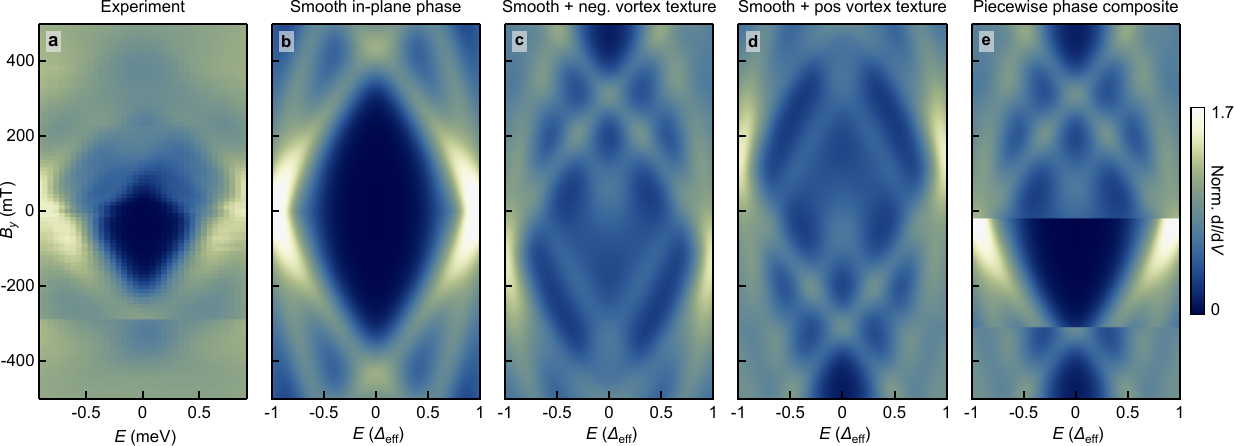}
    \caption{
    \textbf{Phenomenological comparison of abrupt field-asymmetric spectral evolution with added vortex phase textures.}
    Differential-conductance map measured as a function of energy and in-plane magnetic field, compared with trajectory-ensemble simulations using different phenomenological condensate-phase configurations. The data were acquired on the same Cu(111) island as Figs.~3 and 4, at a position near the region where the vortex-induced phase texture is probed in Fig.~4.
    \textbf{a} Experimental differential-conductance map showing abrupt, field-asymmetric spectral rearrangements.
    \textbf{b} Simulation using only the smooth in-plane-field phase accumulation.
    \textbf{c,d} Simulations including the smooth in-plane-field phase accumulation together with an added idealised vortex phase texture of opposite sign.
    \textbf{e} Piecewise phase composite formed from the three calculated phase configurations in \textbf{b--d}: smooth in-plane phase plus negative vortex texture at negative fields up to the first abrupt jump, smooth in-plane phase only between the two jumps, and smooth in-plane phase plus positive vortex texture after the second jump. The comparison is phenomenological: the idealised vortex phase contribution is added to the smooth in-plane phase accumulation as a minimal phase-bias model, without a self-consistent calculation of the superconducting current distribution. The highest intensities in \textbf{b} and \textbf{e} exceed the common colour-scale range and are clipped to preserve visibility of the low-energy features across all panels.
    Parameters: \textbf{a} \(V_\mathrm{stab}{=}500\)\,mV, \(I_\mathrm{stab}{=}500\)\,nA, \(V_\mathrm{mod}{=}50\)\,µV.
    }
    \label{fig:ED_PhaseJump}
\end{figure*}

\end{document}


\maketitle

\section{Trajectory-ensemble model and simulation details}\label{sec:SI_Model}

The trajectory-ensemble model used here is a semiclassical approximation to the full two-dimensional Bogoliubov--de Gennes problem of the proximitised Cu(111) surface state. In a full scattering description, the subgap spectrum would be obtained from a multi-channel Andreev condition of the form
\begin{equation}
\det\left[
1
-
S_h(E)\,r_A(E)\,S_e(E)\,r_A(E)
\right]
=0 .
\label{eq:SI_scattering_determinant}
\end{equation}
Here \(S_e(E)\) and \(S_h(E)\) are the normal-state electron and hole scattering matrices of the two-dimensional island, and \(r_A(E)\) contains the Andreev reflection amplitudes and superconducting phases. In general, Eq.~\eqref{eq:SI_scattering_determinant} mixes transverse modes and different rays.

The present model retains only the diagonal semiclassical contribution of each ray \(n\). Each valid ray is therefore treated as an effective one-dimensional Andreev trajectory between two boundary conversion points, with length \(\ell_n\), transparency \(T_n\), and trajectory-dependent phase bias \(\delta\varphi_n\). In the diagonal approximation,
\begin{align*}
S_e(E) &\rightarrow \mathrm{diag}\left[\exp\left(i k_e \ell_n\right)\right], \\
S_h(E) &\rightarrow \mathrm{diag}\left[\exp\left(-i k_h \ell_n\right)\right] .
\label{eq:SI_diagonal_scattering}
\end{align*}
For \(E,\Delta_{\mathrm{eff}}\ll E_{\mathrm F}\), the Andreev approximation gives
\begin{equation*}
k_e-k_h \simeq \frac{2E}{\hbar v_{\mathrm F}} ,
\label{eq:SI_andreev_k_mismatch}
\end{equation*}
so propagation along the trajectory contributes the dynamical phase
\begin{equation*}
\varphi_{\mathrm{dyn},n}(E)
=
-\frac{2E\ell_n}{\hbar v_{\mathrm F}} .
\label{eq:SI_dyn_phase}
\end{equation*}
In the transparent limit, the corresponding single-ray quantisation condition is
\begin{equation*}
2\arccos\left(\frac{E}{\Delta_{\mathrm{eff}}}\right)
-
\frac{2E\ell_n}{\hbar v_{\mathrm F}}
\mp
\delta\varphi_n
=
2\pi m .
\label{eq:SI_transparent_single_ray}
\end{equation*}

The simulations below use the finite-transparency version of this condition, described in Sec.~\ref{sec:SI_interface_transparency}. Coupling between different rays would appear as off-diagonal terms in the multi-channel scattering problem and would lead to hybridisation and spectral-weight redistribution. The calculated spectra should therefore be interpreted as broadened trajectory-ensemble LDOS envelopes, not as assignments of individual calculated branches to exact eigenstates of the full two-dimensional island.

\subsection{Geometry extraction and trajectory sampling}

Island geometries were extracted from background-subtracted STM topographs by generating binary masks of the Cu(111) island area. The mask value is one inside the island and zero outside. All mask coordinates were calibrated in nanometres using the STM image scaling.

For a given spectroscopy position \(\mathbf r_0=(x_0,y_0)\), the trajectory ensemble was constructed by tracing straight chords through \(\mathbf r_0\) at angles
\begin{equation*}
    \alpha_j =
    \left(j+\frac{1}{2}\right)\frac{\pi}{N_\phi},
    \qquad
    j=0,\ldots,N_\phi-1 .
\end{equation*}
The angular sampling was chosen from the island area \(A\) and the Cu(111) surface-state Fermi wavelength \(\lambda_\mathrm{F}\)
\begin{equation*}
    N_\phi =
    \mathrm{round}
    \left(
        \frac{\pi\sqrt{A}}{\lambda_\mathrm{F}}
    \right),
\end{equation*}
with a minimum of one trajectory. For each angle, the code marches in both directions from \(\mathbf r_0\) until the mask boundary is reached, yielding two boundary intersection points
\begin{equation*}
    \mathbf r_{1,n}=(x_{1,n},y_{1,n}),
    \qquad
    \mathbf r_{2,n}=(x_{2,n},y_{2,n}) .
\end{equation*}
Only trajectories for which both boundary intersections are found are retained.

For each valid trajectory \(n\), the chord length is
\begin{equation*}
    \ell_n =
    \left|
        \mathbf r_{2,n}-\mathbf r_{1,n}
    \right|,
\end{equation*}
and the magnetic lever arm for an in-plane field direction \(\hat{\mathbf e}_B\) is
\begin{equation*}
    w_n =
    \left|
    \left(
        \mathbf r_{2,n}-\mathbf r_{1,n}
    \right)
    \cdot
    \hat{\mathbf e}_{\perp}
    \right|,
    \qquad
    \hat{\mathbf e}_{\perp}\perp \hat{\mathbf e}_B .
\end{equation*}
Thus \(\ell_n\) controls the dynamical phase accumulated along the trajectory, whereas \(w_n\) controls the in-plane-field phase accumulation.

The trajectory weights were taken to be equal for all valid two-dimensional trajectories and normalised such that
\[
    \sum_n a_n = 1 .
\]
Bundle-derived values of \(\ell\) and \(w\) were extracted from the same trajectory ensembles, as described below.

\subsection{Extraction of bundle-derived geometric scales}

The LDOS simulations use the full trajectory ensemble rather than a single selected trajectory. For the main-text scaling plots, we additionally extract bundle-derived geometric scales from the same ensemble. They identify sufficiently populated trajectory bundles that set the dominant phase-accumulation scales used for comparison with the experimentally extracted \(E_0\) and \(B_0\).

For each valid trajectory \(n\), the individual trajectory length and transverse projection are denoted by \(\ell_n\) and \(w_n\), respectively. The unindexed quantities \(\ell\) and \(w\) denote bundle-derived scales extracted from the corresponding distributions. For a generic trajectory quantity
\[
x_n \in \{\ell_n,w_n\},
\]
the values are grouped into bins of width
\[
\Delta x = 20\,\mathrm{nm}.
\]

The bin width was chosen as an order-of-magnitude finite-coherence coarse-graining scale. A finite phase-coherence length \(L_\phi\) defines a small energy scale
\[
\delta E_\phi \sim \frac{\hbar v_\mathrm{F}}{L_\phi}.
\]
It provides a practical estimate for when nearby trajectory families should no longer be distinguished as independent geometric scales. Comparing \(\delta E_\phi\) with the superconducting scale \(\Delta_{\mathrm{eff}}\) over the effective Andreev length
\[
\xi_{\mathrm{eff}}\sim\frac{\hbar v_\mathrm{F}}{\Delta_{\mathrm{eff}}},
\]
gives the associated geometric scale
\[
\delta \ell
\sim
\xi_{\mathrm{eff}}
\frac{\delta E_\phi}{\Delta_{\mathrm{eff}}}
=
\frac{(\hbar v_\mathrm{F})^2}
{L_\phi\Delta_{\mathrm{eff}}^2}.
\]
With \(v_\mathrm{F}=5.8\times10^5\,\mathrm{m\,s^{-1}}\),
\(\Delta_{\mathrm{eff}}=0.92\,\mathrm{meV}\), and
\(L_\phi\simeq10\,\text{µ}\mathrm{m}\), this gives
\[
\xi_{\mathrm{eff}}\simeq 415\,\mathrm{nm},
\qquad
\delta \ell \simeq 17\,\mathrm{nm}.
\]
We therefore use \(20\,\mathrm{nm}\) bins for both \(\ell_n\) and \(w_n\). The bin width should be understood as a practical finite-coherence coarse graining for extracting trajectory bundles, not as a fundamental spatial resolution or as part of the ray sampling itself.

Only bins containing at least
\[
N_{\min}=\max\left(5,\mathrm{round}(0.01N)\right)
\]
trajectories are retained as trajectory bundles, where \(N\) is the total number of valid trajectories in the ensemble. This minimum-population criterion suppresses isolated extremal rays and small bundles that are not robust against mask thresholding, boundary discretisation and finite angular sampling. For the zero-field spectral scale, we select the retained bundle with the largest \(\ell_n\) values. For the magnetic-field scale, we select the retained bundle with the largest \(w_n\) values for the corresponding field direction. If several retained bundles satisfy the criterion, the bundle with the largest upper-bin edge is used.

The bundle-derived scale is defined as the maximum trajectory value within the selected bundle,
\[
\ell \equiv \max_{n\in\mathrm{bundle}} \ell_n,
\qquad
w \equiv \max_{n\in\mathrm{bundle}} w_n .
\]
Displayed trajectories are the ray-traced trajectories closest to the corresponding bundle-derived values. Bundle medians, widths and populations were stored as diagnostics and used to test alternative scale definitions. These alternatives gave the same qualitative scaling trends. We use the maximum value within the selected bundle because it is less sensitive to the precise binning convention and remained stable for reasonable variations of the bin width.

\subsection{Interface transparency and finite-transparency ABS condition}
\label{sec:SI_interface_transparency}
Finite normal reflection at the island boundary was included through a BTK-like dimensionless barrier parameter \(Z\). At each boundary hit, the local incidence angle \(\theta\) was determined from the ray direction and the local mask-boundary normal. The local transparency was then taken as
\begin{equation*}
    T(\theta)
    =
    \frac{\cos^2\theta}{\cos^2\theta+Z^2}.
\end{equation*}
For a trajectory connecting two boundary points, the effective transparency was chosen as
\begin{equation*}
    T_n=\min(T_{1,n},T_{2,n}),
\end{equation*}
where \(T_{1,n}\) and \(T_{2,n}\) are the transparencies at the two boundary intersections. Unless stated otherwise, we used
\begin{equation*}
    Z=0.05,
\end{equation*}
corresponding to a highly transparent boundary between the Cu(111) surface state and the superconductor.

For each trajectory, we define
\begin{equation*}
    \Phi_n(E)
    =
    2\arccos
    \left(
        \frac{E}{\Delta_{\mathrm{eff}}}
    \right)
    -
    \frac{2E\ell_n}{\hbar v_\mathrm{F}},
\end{equation*}
where \(\Delta_{\mathrm{eff}}\) is the effective superconducting gap scale experienced by the Cu(111) surface state and \(v_\mathrm{F}\) is the Cu(111) surface-state Fermi velocity. The finite-transparency condition is
\begin{equation}
    \cos\Phi_n(E)
    =
    (1-T_n)+T_n\cos\!\left(\delta\varphi_n\right) .
    \label{eq:SI_finiteT}
\end{equation}
Here \(\delta\varphi_n\) is the trajectory-dependent phase bias entering the bound-state condition. It is the trajectory-resolved counterpart of the gauge-invariant phase bias introduced in the main text. Equivalently, defining
\begin{equation*}
    C_n=(1-T_n)+T_n\cos\!\left(\delta\varphi_n\right),
    \qquad
    \theta_n=\arccos C_n,
\end{equation*}
the branches satisfy
\begin{equation*}
    \Phi_n(E)=\pm \theta_n+2\pi m,
    \qquad
    m\in\mathbb Z .
\end{equation*}
The numerical solver searches the allowed energy range \(|E|<\Delta_{\mathrm{eff}}\) for all roots of Eq.~\eqref{eq:SI_finiteT}. Arguments of inverse trigonometric functions are clipped to their physical interval to avoid numerical artefacts at the gap edge.

In the transparent limit \(T_n\rightarrow1\), Eq.~\eqref{eq:SI_finiteT} reduces to the quantisation condition
\begin{equation*}
    2\arccos
    \left(
        \frac{E}{\Delta_{\mathrm{eff}}}
    \right)
    -
    \frac{2E\ell_n}{\hbar v_\mathrm{F}}
    \mp \delta\varphi_n
    =
    2\pi m
\end{equation*}
used in the main text. At \(E=0\), the dynamical term vanishes, so branches with similar magnetic lever arm reach zero energy near the same odd-\(\pi\) phase condition even if their zero-field energies differ. 

\subsection{Magnetic-field-induced phase from in-plane magnetic fields}

For an in-plane magnetic field without additional phase textures, we used
\begin{equation}
    \left|\delta\varphi_{B,n}\right|
    =
    \frac{2\pi}{\Phi_0}
    |B| h_{\mathrm{eff}} w_n,
    \label{eq:SI_betaB}
\end{equation}
where \(\Phi_0=h/2e\). The parameter \(h_{\mathrm{eff}}\) is the effective vertical phase lever arm that converts the transverse projection \(w_n\) into the effective area \(h_{\mathrm{eff}}w_n\) for in-plane-field phase accumulation. Unless stated otherwise
\begin{equation*}
    h_{\mathrm{eff}} = 24~\mathrm{nm} .
\end{equation*}
The sign of the trajectory phase depends on the field and trajectory orientation convention. The calculated density of states is invariant under the corresponding branch relabelling.

For the lowest positive zero-bias crossing in the transparent limit, Eq.~\eqref{eq:SI_betaB} gives
\begin{equation*}
    B_0
    \simeq
    \frac{\Phi_0}{2h_{\mathrm{eff}}w_n}.
\end{equation*}
Higher unfolded branch crossings occur at odd multiples of this phase condition.

\subsection{Additional phase textures}

Additional superconducting phase textures were included phenomenologically through a trajectory-dependent phase term
\[
    \delta\varphi_n(B)
    =
    \delta\varphi_{B,n}(B)
    +
    \delta\varphi_{\mathrm{extra},n}(B).
\]
This provides a minimal way to test how the trajectory ensemble responds to prescribed condensate phase textures without solving the superconducting current distribution self-consistently.

For idealised free-vortex simulations, the phase field was centred at \((x_{\mathrm v},y_{\mathrm v})\),
\[
    \phi_{\mathrm v}(\mathbf r)
    =
    \operatorname{atan2}
    \left(
        y-y_{\mathrm v},
        x-x_{\mathrm v}
    \right).
\]
The vortex contribution for trajectory \(n\) was the wrapped phase difference between its boundary endpoints
\[
    \delta\varphi_{\mathrm v,n}
    =
    n_\Phi\,
    \mathrm{Arg}
    \left[
        \exp
        \left\{
        \mathrm{i}\left[
            \phi_{\mathrm v}(\mathbf r_{2,n})
            -
            \phi_{\mathrm v}(\mathbf r_{1,n})
        \right]
        \right\}
    \right],
\]
where \(n_\Phi\) is the vorticity. These calculations are qualitative comparisons of different local phase configurations, not self-consistent vortex-entry simulations.

\subsection{Parameter values and LDOS construction}

The Cu(111) surface-state parameters were obtained from large-bias spectroscopy of the surface-state onset and the parabolic dispersion. Unless stated otherwise, the simulations used
\begin{equation*}
\begin{aligned}
    E_\mathrm{F} &= 0.429~\mathrm{eV}, &
    m^\ast &= 0.45m_e, \\
    \Delta_{\mathrm{eff}} &= 0.92~\mathrm{meV}, &
    h_{\mathrm{eff}} &= 24~\mathrm{nm}, \\
    Z &= 0.05 .
\end{aligned}
\end{equation*}
These values give
\begin{equation*}
    v_\mathrm{F}
    =
    \sqrt{\frac{2E_\mathrm{F}}{m^\ast}}
    \simeq
    5.8\times10^5~\mathrm{m\,s^{-1}},
\end{equation*}
and
\begin{equation*}
    \lambda_\mathrm{F}
    =
    \frac{2\pi\hbar}{m^\ast v_\mathrm{F}}
    \simeq
    2.8~\mathrm{nm}.
\end{equation*}

For each trajectory and magnetic-field value, the solver returns a set of branch energies \(E_{n,j}(B)\). The simulated local density of states was constructed by summing Lorentzian contributions from all branches and all trajectories,
\begin{equation*}
    \rho(E,B)
    =
    \sum_n a_n
    \sum_j
    \frac{1}{\pi}
    \frac{\Gamma}
    {
        \left[E-E_{n,j}(B)\right]^2+\Gamma^2
    } .
\end{equation*}
A Lorentzian linewidth \(\Gamma\) was applied before convolution with the experimental energy resolution. Thermal and lock-in broadening were applied by convolution with the corresponding resolution kernels using the effective electronic temperature and modulation voltage. Additional field-dependent broadening terms were included only where explicitly stated.

\subsection{Bulk-derived three-dimensional trajectory sampling}

For the bulk-derived channel, a three-dimensional trajectory ensemble was generated from the same STM-derived island mask. The mask was interpreted as the top-surface outline at \(z=H\), with the superconducting interface at \(z=0\). The island volume was approximated by a Cu(111)-consistent faceted prism. The lower cross-section was expanded laterally according to
\[
    \frac{\Delta r_{\parallel}}{\Delta z}=\frac{1}{\sqrt{2}},
\]
corresponding to a side-wall angle of about \(35^\circ\) from the vertical.

For a spectroscopy position \((x_0,y_0,H)\), rays were launched into the island volume over downward directions. For each direction, two legs with opposite in-plane components were traced to the superconducting plane, rather than treating the channel as a single chord, because the three-dimensional trajectories need not remain collinear and may include side-wall reflections. A trajectory was retained only if both legs reached valid points on the \(z=0\) interface inside the faceted prism. Specular reflections at the side walls were included when a leg reached the vacuum boundary before the superconducting interface.

For each valid trajectory, the total length was
\[
    \ell_n=\ell_{1,n}+\ell_{2,n}.
\]
The geometric transverse projection was calculated from the two interface points
\[
    w_{\mathrm{geom},n}
    =
    \left|
    \left(
        \mathbf r_{2,n}-\mathbf r_{1,n}
    \right)
    \cdot
    \hat{\mathbf e}_{\perp}
    \right|.
\]
For the in-plane-field phase, the tracer evaluated the signed effective magnetic area \(A^{\mathrm{3D}}_{\mathrm{eff},n}\), defined from the trajectory integral of the vector potential for a unit in-plane field expressed as an equivalent flux area. Its sign tracks the orientation of the trajectory with respect to the field. This was converted into a solver-equivalent width
\[
    w_{\mathrm{eff},n}
    =
    \frac{|A^{\mathrm{3D}}_{\mathrm{eff},n}|}{h_{\mathrm{eff}}},
\]
so that the same phase form as in the two-dimensional solver could be used,
\[
    \left|\delta\varphi_{B,n}^{\mathrm{3D}}\right|
    =
    \frac{2\pi}{\Phi_0}
    |B|h_{\mathrm{eff}}w_{\mathrm{eff},n}.
\]

The three-dimensional channel transparency was evaluated from the incidence angles at the two superconducting-interface hits, and the channel transparency was taken as \(T_n=\min(T_{1,n},T_{2,n})\). Angular weights \(a_n\propto\sin\theta_n\) were normalised during LDOS construction.

For the bulk-derived spectral calculation, the same finite-transparency solver was used with bulk-like parameters
\[
    E_{\mathrm{F},\mathrm{3D}}=7~\mathrm{eV},
    \qquad
    m^\ast_{\mathrm{3D}}=m_e,
    \qquad
    \Delta_{\mathrm{3D}}=1.3~\mathrm{meV}.
\]
These values give
\[
    v_{\mathrm{F},\mathrm{3D}}
    =
    \sqrt{
        \frac{2E_{\mathrm{F},\mathrm{3D}}}{m_e}
    }
    \simeq
    1.57\times10^6~\mathrm{m\,s^{-1}} .
\]
The three-dimensional LDOS was computed from the same Lorentzian branch summation as the two-dimensional LDOS, using \(\ell_n\), \(w_{\mathrm{eff},n}\), \(T_n\), and \(a_n\).

\section{Large-bias determination of the surface-state onset energy}

The Cu(111) surface-state onset energy was determined from large-bias line spectroscopy on island A. For each spectrum along a line, the onset was extracted by fitting the step-like increase in \(\mathrm{d}I/\mathrm{d}V\). The fitted local onset energies were collected in a histogram, and their mean value was used as the surface-state band bottom relative to the Fermi level.

\begin{figure}[!ht]
    \centering
    \includegraphics[width=0.6\linewidth]{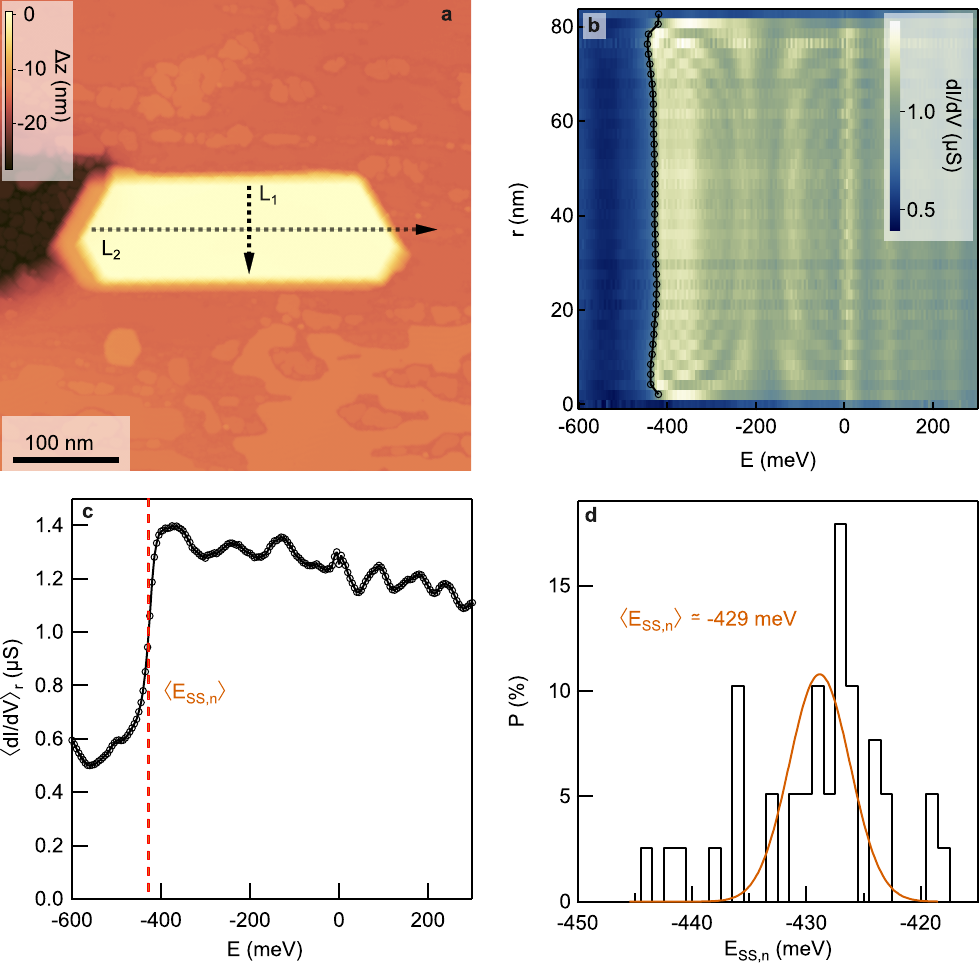}
    \caption{
    \textbf{Large-bias determination of the Cu(111) surface-state onset energy.}
    \textbf{a} STM topography of island A from Fig.~1, with the two large-bias line-spectroscopy paths used for the onset analysis indicated. Here, \(L_1\) denotes the short-axis path and \(L_2\) the long-axis path.
    \textbf{b} Representative large-bias \(\mathrm{d}I/\mathrm{d}V\) map measured along \(L_1\). Black markers indicate fitted local onset energies \(E_{\mathrm{SS},n}\) for spectra with valid fits.
    \textbf{c} Spatially averaged large-bias spectrum \(\langle \mathrm{d}I/\mathrm{d}V\rangle_r\), obtained from the line-spectroscopy data, showing the step-like onset of the Cu(111) surface-state band. The red dashed line marks the mean fitted onset energy \(\langle E_{\mathrm{SS},n}\rangle\).
    \textbf{d} Histogram of fitted local onset energies \(E_{\mathrm{SS},n}\) from both line-spectroscopy data sets using \(1~\mathrm{meV}\) bins. A Gaussian fit gives \(\langle E_{\mathrm{SS},n}\rangle\simeq -429~\mathrm{meV}\) with \(\sigma\simeq4~\mathrm{meV}\), corresponding to a band bottom \(429~\mathrm{meV}\) below the Fermi level.
    Parameters: \textbf{a} \(V{=}900\)\,mV, \(I{=}1.1\)\,nA;
    \textbf{b} \(V{=}500\)\,mV, \(I{=}500\)\,nA, \(V_\mathrm{mod}{=}5\)\,mV.
    }
    \label{fig:SI_SurfaceStateEnergy_island7}
\end{figure}

\section{Axis correction and extraction of experimental spectral scales}

This section describes the analysis used to extract the quantities entering the main-text scaling plots. The bundle-derived trajectory length \(\ell\) and transverse projection \(w\) were obtained by applying the ray-tracing and bundle-extraction procedure of Section~\ref{sec:SI_Model} to STM-derived island masks. The spectral scales \(E_0\) and \(B_0\) were extracted from the corresponding field-dependent tunnelling spectra as described below.

\begin{table}[!ht]
\centering
\caption{
Summary of island parameters used in the scaling analysis.
Internal IDs refer to the original data-set numbering.
The spectral scales \(E_0\) and \(B_0\) were extracted from corrected tunnelling spectra, with uncertainties defined below.
The bundle-derived trajectory length \(\ell\) and transverse projection \(w\) were obtained from ray-traced ensembles of STM-derived island masks. 10\% uncertainties were used in the scaling plots to account for calibration and drift in the topography measurements.
\\
}
\begin{tabular}{lllccccc}
\hline\hline
Island & Internal ID & Field direction
& \(h_{\mathrm{Cu}}\) (nm)
& \(E_0\) (meV)
& \(B_0\) (mT)
& \(\ell\) (nm)
& \(w\) (nm) \\
\hline
A & I7, Set1 & \(B_y\)          & 15 & \(0.78 \pm 0.13\) & \(154 \pm 20\) & 274 & 273 \\
A & I7, Set3 & \(B_{45^\circ}\) & 15 & \(0.77 \pm 0.15\) & \(288 \pm 20\) & 274 & 228 \\
B & I3, Set1 & \(B_y\)          & 14 & \(0.81 \pm 0.09\) & \(268 \pm 20\) & 214 & 197 \\
C & I4, Set1 & \(B_y\)          & 15 & \(0.85 \pm 0.13\) & \(162 \pm 20\) & 287 & 286 \\
C & I4, Set2 & \(B_x\)          & 15 & \(0.84 \pm 0.14\) & \(457 \pm 60\) & 287 & 92  \\
D & I1, Set3 & \(B_y\)          & 14 & \(0.63 \pm 0.15\) & \(63 \pm 10\)  & 624 & 624 \\
D & I1, Set2 & \(B_x\)          & 14 & \(0.62 \pm 0.10\) & \(199 \pm 14\) & 624 & 158 \\
E & I2, Set1 & \(B_y\)          & 19 & \(0.89 \pm 0.14\) & \(229 \pm 20\) & 253 & 250 \\
F & I5, Set1 & \(B_y\)          & 11 & \(0.67 \pm 0.09\) & \(120 \pm 20\) & 370 & 332 \\
G & I6, Set1 & \(B_y\)          & 32 & \(0.81 \pm 0.11\) & \(208 \pm 20\) & 175 & 167 \\
\hline\hline
\end{tabular}
\label{tab:island_overview}
\end{table}

\FloatBarrier

\subsection{Correction of magnetic-field hysteresis}
\label{sec:SM_hysteresis_correction}

Magnetic-field hysteresis produces a small difference between the nominal field and the effective field experienced by the sample. This leads to a sweep-direction-dependent displacement of the conductance pattern along the field axis, as shown in Supplementary Fig.~\ref{fig:SM_hysteresis_correction}. For each sweep, we determined the symmetry centre of the low-field subgap pattern from the normalised, interpolated conductance map \((\mathrm{d}I/\mathrm{d}V)_{\mathrm{norm}}(E_{\mathrm{set}},B_{\mathrm{set}})\), using its approximate mirror symmetry about \(E=0\) and \(B=0\). Nearly flat regions, field-range edges, abrupt jumps and weakly structured background were excluded. The resulting correction offsets \(E_{\mathrm{corr}}\) and \(B_{\mathrm{corr}}\) were applied as
\[
E=E_{\mathrm{set}}+E_{\mathrm{corr}}, \qquad
B=B_{\mathrm{set}}+B_{\mathrm{corr}} .
\]
All zero-field and zero-bias cuts used to extract \(E_0\) and \(B_0\) were taken from these axis-corrected maps.

\begin{figure}[h!]
    \centering
    \includegraphics[width=\textwidth]{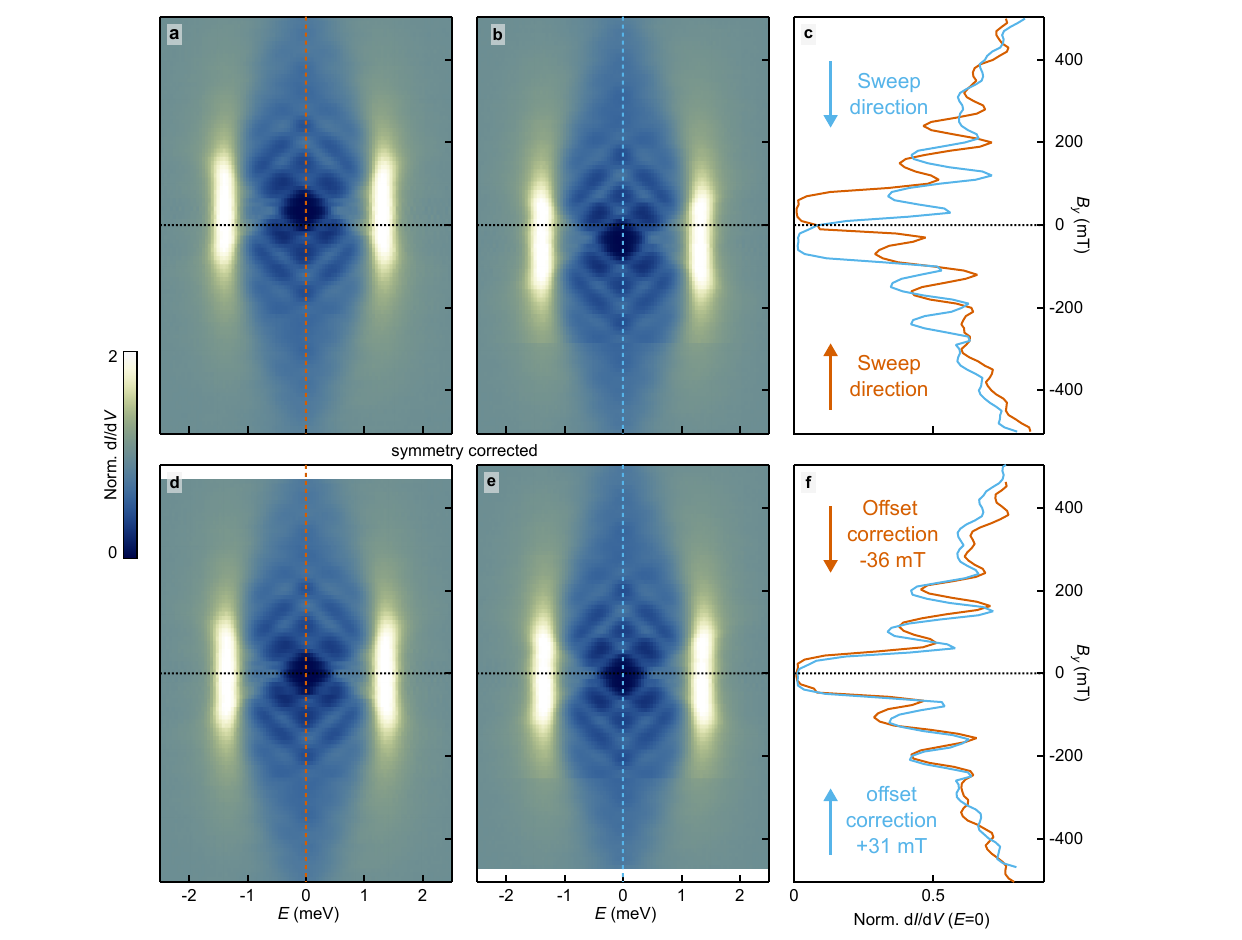}
    \caption{
    \textbf{Correction of magnetic-field hysteresis.}
    \textbf{a,b} Normalised \(\mathrm{d}I/\mathrm{d}V\) maps recorded at the same STS position on island D for opposite sweep directions. The dashed orange line marks the zero-bias cut shown in \textbf{c}; the black dotted line marks the nominal zero-field axis.
    \textbf{c} Zero-bias cuts \((\mathrm{d}I/\mathrm{d}V)_{\mathrm{norm}}(E=0,B)\) before correction, showing the sweep-direction-dependent field offset.
    \textbf{d,e} Same maps after axis correction using the symmetry centre of the subgap pattern.
    \textbf{f} Zero-bias cuts after correction. Field-axis shifts of \(B_\mathrm{corr}\approx-36~\mathrm{mT}\) and \(B_\mathrm{corr}\approx+31~\mathrm{mT}\) align the low-field conductance features. Corrections on the energy axis are minor with \(E_\mathrm{corr}\approx+45\)\,µeV and \(E_\mathrm{corr}\approx+38\)\,µeV, respectively.
    Parameters: \textbf{a--f} \(V_\mathrm{stab}{=}500\)\,mV, \(I_\mathrm{stab}{=}500\)\,nA, \(V_\mathrm{mod}{=}50\)\,µV.
    }
    \label{fig:SM_hysteresis_correction}
\end{figure}

\FloatBarrier

\subsection{Extraction of \texorpdfstring{\(E_{0}\)}{E0} and \texorpdfstring{\(B_{0}\)}{B0} from corrected conductance maps}
\label{sec:SI_E0_B0_extraction}

For each island and field direction, the field-dependent tunnelling spectra were assembled into normalised conductance maps \((\mathrm{d}I/\mathrm{d}V)_{\mathrm{norm}}(E,B)\). Before extracting \(E_0\) and \(B_0\), the spectra were normalised to the above-gap conductance, interpolated in energy and field, and corrected for the zero-bias and zero-field offsets described above.

The characteristic field scale \(B_0\) was determined from the zero-bias cut \((\mathrm{d}I/\mathrm{d}V)_{\mathrm{norm}}(E=0,B)\). Local maxima in this trace identify fields at which the lowest-energy Andreev feature approaches zero bias. If two approximately symmetric maxima were resolved at \(B_-<0\) and \(B_+>0\), we defined
\[
    B_0 = \frac{B_+-B_-}{2}.
\]
For sweeps in which only one side of the gap closing was cleanly resolved, we used
\[
    B_0 = |B_{\mathrm{pk}}|,
\]
where \(B_{\mathrm{pk}}\) is the selected zero-bias maximum. The uncertainty in \(B_0\) was estimated from the field-axis resolution and the ambiguity in selecting the zero-bias maximum. In the figures, dashed horizontal lines mark the selected maxima, and arrows indicate either \(2B_0\) for paired extractions or \(B_0\) for single-sided extractions.

The zero-field excitation energy \(E_0\) was extracted from
\((\mathrm{d}I/\mathrm{d}V)_{\mathrm{norm}}(E,B=0)\). Because several Andreev levels contribute within the low-energy window and overlap once linewidth and experimental energy resolution are included, \(E_0\) was not assigned from a single peak position. Instead, the relevant spectral feature was bracketed by a lower and an upper energy bound on each side of zero bias.

The lower bound was taken from the maxima in the absolute first derivative
\[
    \left|
    \frac{\partial}{\partial E}
    \left(\frac{\mathrm{d}I}{\mathrm{d}V}\right)_{\mathrm{norm}}(E,B=0)
    \right|,
\]
which identify the steepest rising edges of the low-energy feature. The upper bound was taken from the corresponding local maxima in the zero-field conductance spectrum.

For the negative- and positive-energy lower bounds \(E_-^{\mathrm{low}}\) and \(E_+^{\mathrm{low}}\), and the corresponding upper bounds \(E_-^{\mathrm{high}}\) and \(E_+^{\mathrm{high}}\), we defined
\[
    E_0^{\mathrm{low}} =
    \frac{E_+^{\mathrm{low}}-E_-^{\mathrm{low}}}{2},
    \qquad
    E_0^{\mathrm{high}} =
    \frac{E_+^{\mathrm{high}}-E_-^{\mathrm{high}}}{2}.
\]
The reported value is
\[
    E_0 =
    \frac{E_0^{\mathrm{low}}+E_0^{\mathrm{high}}}{2},
\]
with uncertainty
\[
    \delta E_0 =
    \frac{\left|E_0^{\mathrm{high}}-E_0^{\mathrm{low}}\right|}{2}.
\]
This uncertainty reflects the finite linewidth, experimental energy resolution and overlap of several Andreev levels in the low-energy spectral feature.
In the zero-field panels, dashed vertical lines and shaded intervals indicate the bounds used for this procedure, and arrows labelled \(2E_0\) connect the centres of the negative- and positive-energy brackets.

The topographic panels show the analysed island, spectroscopy position and applied field direction.
The Cu thickness \(h_{\mathrm{Cu}}\) was determined from the height histogram.
Black lines indicate trajectories associated with the bundle-derived \(\ell\) and \(w\) scales.
The extraction of \(E_0\) and \(B_0\) uses only the one-dimensional cuts described above.

\begin{figure}[!ht]
    \centering
    \includegraphics[width=\textwidth]{\detokenize{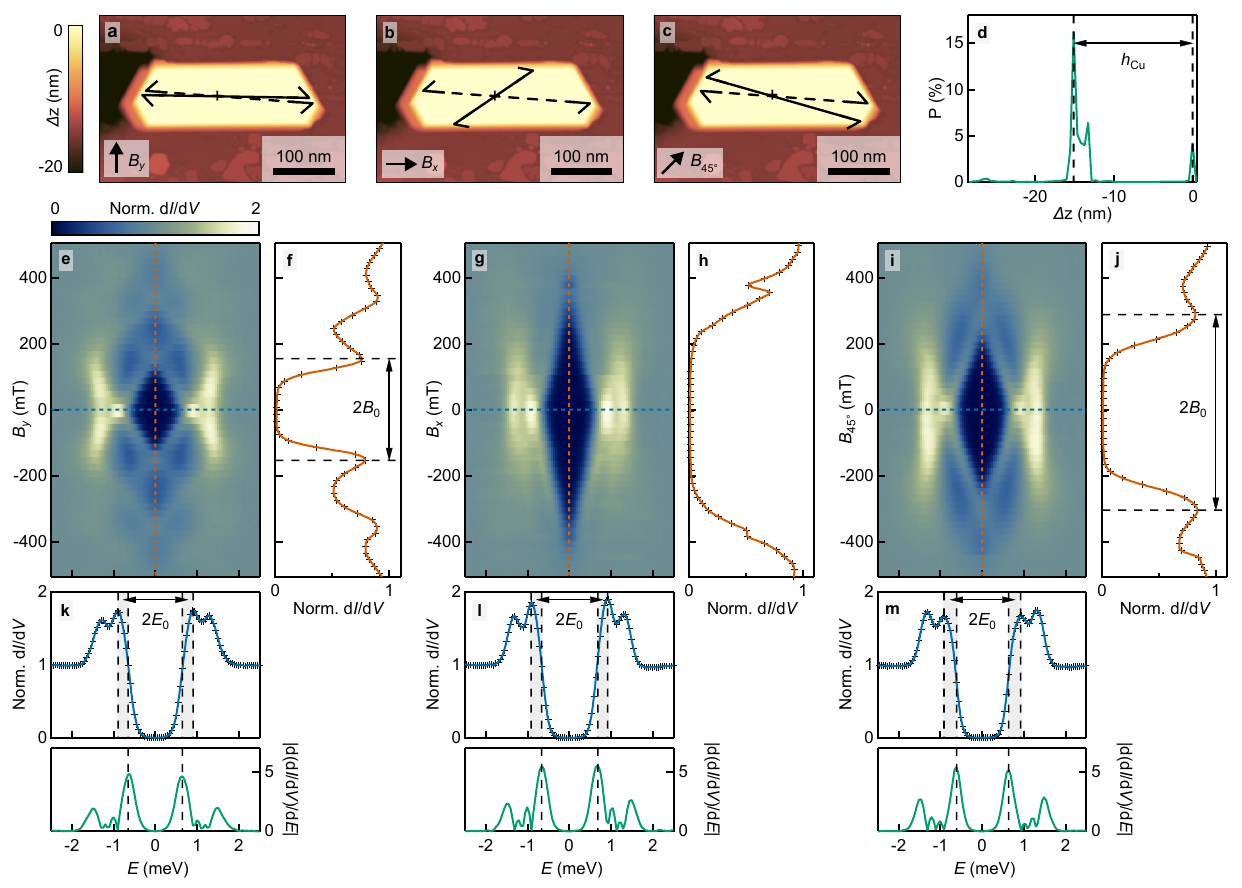}}
    \caption{\textbf{Extraction of \(E_{0}\) and \(B_{0}\) for island A.}
    Island A is shown in Fig.~1.
    \textbf{a--c} Topographic images for \(B_y\), \(B_x\) and \(B_{45^\circ}\). Crosses mark the spectroscopy positions and black lines indicate trajectories associated with the bundle-derived \(\ell\) and \(w\) scales.
    \textbf{d} Height histogram used to determine \(h_{\mathrm{Cu}}\).
    \textbf{e,g,i} Normalised \(\mathrm{d}I/\mathrm{d}V\) maps for the same field directions. Dashed lines mark the zero-bias and zero-field cuts.
    \textbf{f,h,j} Zero-bias conductance cuts. The \(B_y\) and \(B_{45^\circ}\) sweeps show paired maxima used to extract \(2B_0\). The \(B_x\) trace is shown for completeness; jumps in the zero-bias conductance prevent a robust \(B_0\) extraction.
    \textbf{k--m} Zero-field spectra and absolute first derivatives for the same field directions. Dashed vertical lines and shaded brackets define \(E_0^{\mathrm{low}}\) and \(E_0^{\mathrm{high}}\); arrows indicate \(2E_0\).
    Parameters: \textbf{a--c} \(V{=}900\)\,mV, \(I{=}1.1\)\,nA;
    \textbf{e--m} \(V_\mathrm{stab}{=}500\)\,mV, \(I_\mathrm{stab}{=}500\)\,nA, \(V_\mathrm{mod}{=}50\)\,µV.
    }
    \label{fig:SM_Isl7_E0_B0}
\end{figure}

\begin{figure}[!ht]
    \centering
    \includegraphics[width=\textwidth]{\detokenize{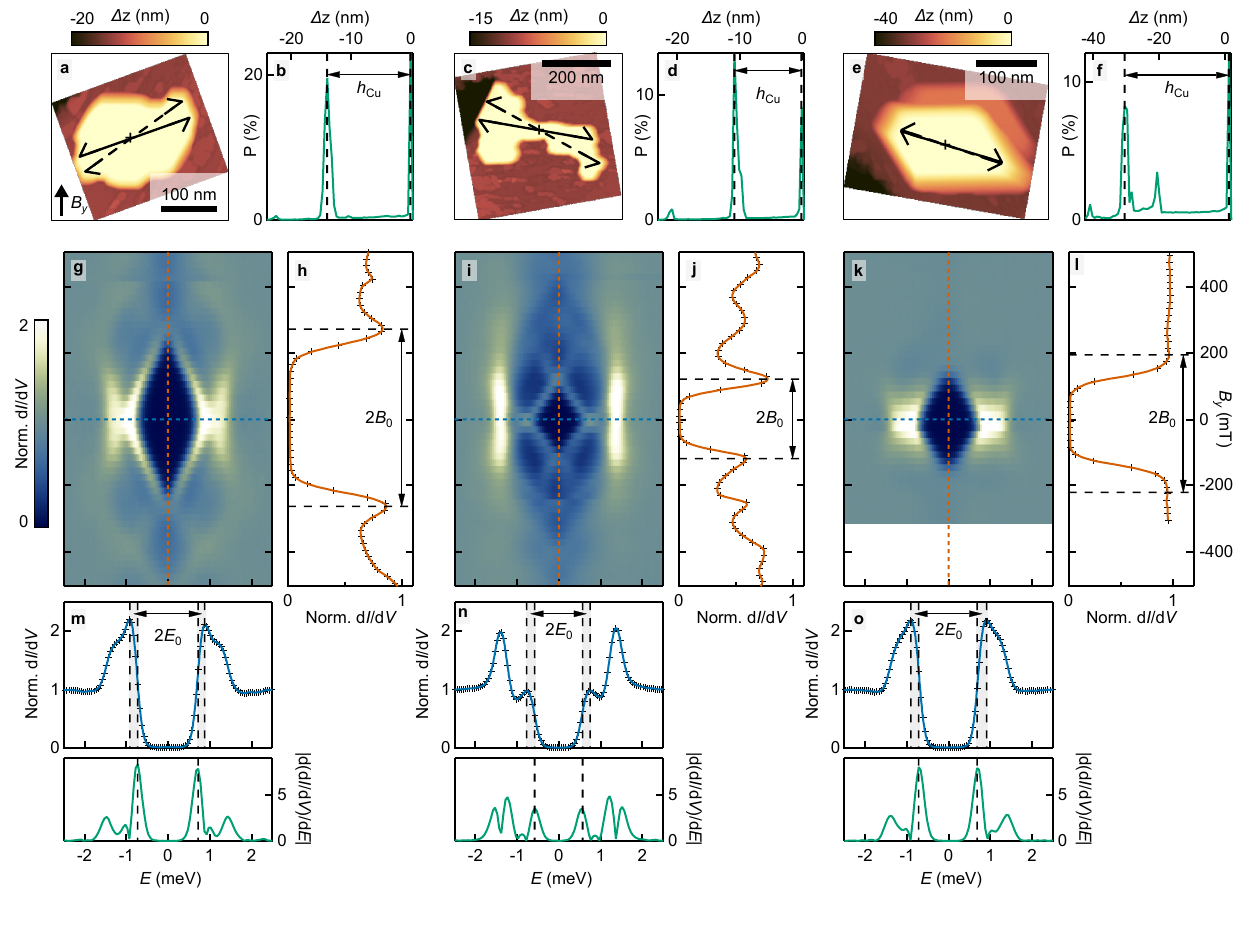}}
    \caption{\textbf{Extraction of \(E_{0}\) and \(B_{0}\) for islands B, F and G.}
    Island B is shown in Fig.~2a.
    The three columns correspond to islands B, F and G, respectively; all sweeps were measured for \(B_y\).
    \textbf{a,c,e} Topographic images showing the spectroscopy positions and trajectories associated with the bundle-derived \(\ell\) and \(w\) scales.
    \textbf{b,d,f} Height histograms used to determine \(h_{\mathrm{Cu}}\).
    \textbf{g,i,k} Normalised \(\mathrm{d}I/\mathrm{d}V\) maps. Dashed lines mark the zero-bias and zero-field cuts.
    \textbf{h,j,l} Zero-bias cuts \((\mathrm{d}I/\mathrm{d}V)_{\mathrm{norm}}(E=0,B)\) used to determine \(B_0\). Dashed horizontal lines mark the selected zero-bias maxima; arrows indicate \(2B_0\).
    \textbf{m,n,o} Zero-field spectra \((\mathrm{d}I/\mathrm{d}V)_{\mathrm{norm}}(E,B=0)\) and absolute first derivatives used to extract \(E_0\). Dashed vertical lines mark the lower and upper energy bounds; arrows denote \(2E_0\).
    Parameters: \textbf{a,e} \(V{=}900\)\,mV, \(I{=}1\)\,nA;
    \textbf{c} \(V{=}500\)\,mV, \(I{=}1.4\)\,nA;
    \textbf{g--o} \(V_\mathrm{stab}{=}500\)\,mV, \(I_\mathrm{stab}{=}500\)\,nA, \(V_\mathrm{mod}{=}50\)\,µV.
    }
    \label{fig:SM_Isl3_5_6_E0_B0}
\end{figure}

\begin{figure}[!ht]
    \centering
    \includegraphics[width=\textwidth]{\detokenize{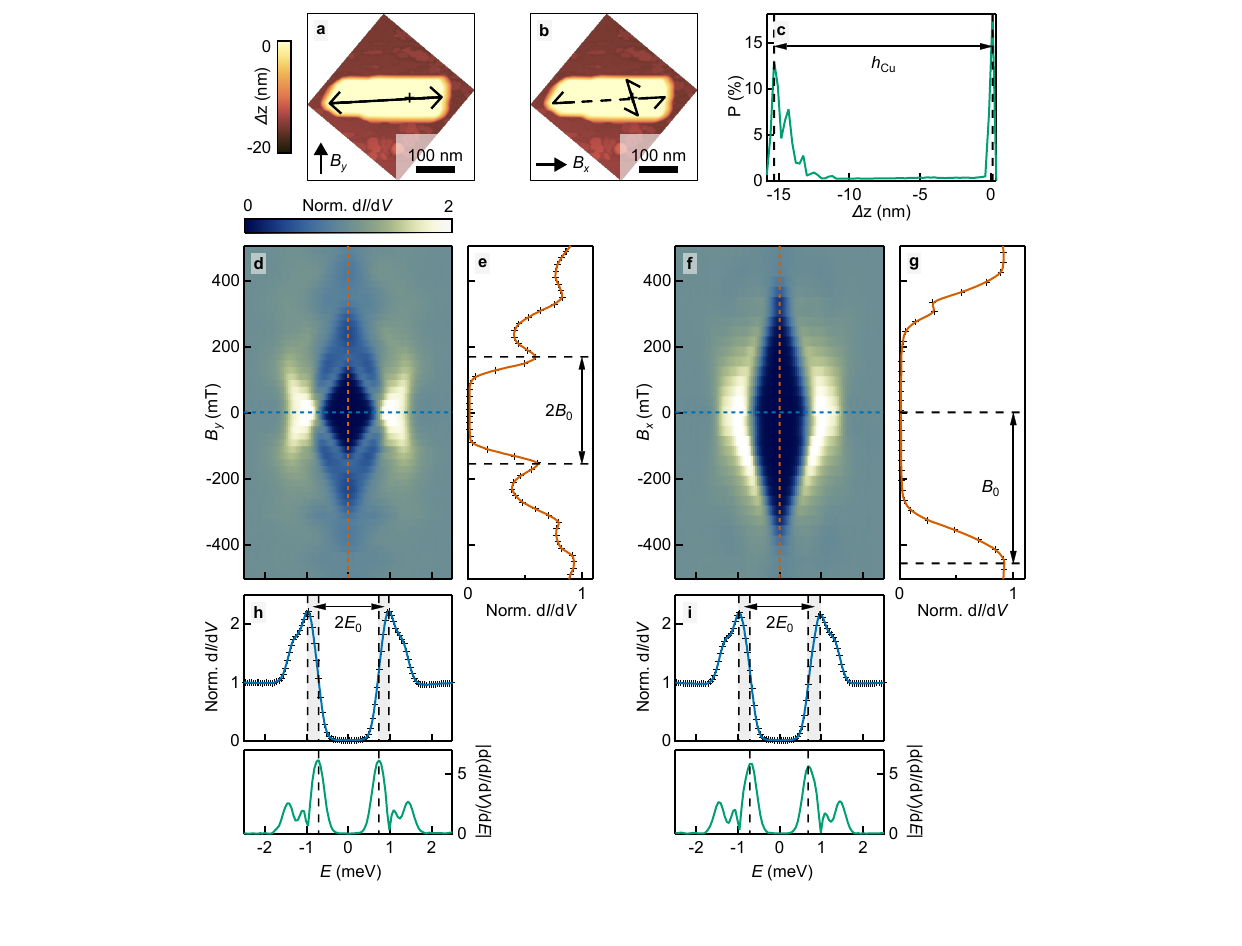}}
    \caption{\textbf{Extraction of \(E_{0}\) and \(B_{0}\) for island C.}
    Island C is shown in Fig.~2b.
    \textbf{a,b} Topographic images for \(B_y\) and \(B_x\). Crosses mark the spectroscopy positions and black lines indicate trajectories associated with the bundle-derived \(\ell\) and \(w\) scales.
    \textbf{c} Height histogram used to determine \(h_{\mathrm{Cu}}\).
    \textbf{d,f} Normalised \(\mathrm{d}I/\mathrm{d}V\) maps for the same field directions. Dashed lines mark the zero-bias and zero-field cuts.
    \textbf{e,g} Zero-bias conductance cuts. Selected maxima define \(B_0\); arrows denote \(2B_0\) for paired peaks and \(B_0\) for single-sided extraction, as labelled.
    \textbf{h,i} Zero-field spectra and absolute first derivatives. Dashed vertical lines and shaded brackets define \(E_0^{\mathrm{low}}\) and \(E_0^{\mathrm{high}}\); arrows indicate \(2E_0\).
    Parameters: \textbf{a,b} \(V{=}900\)\,mV, \(I{=}1\)\,nA;
    \textbf{d--i} \(V_\mathrm{stab}{=}900\)\,mV, \(I_\mathrm{stab}{=}500\)\,nA, \(V_\mathrm{mod}{=}50\)\,µV.
    }
\label{fig:SM_Isl4_E0_B0}
\end{figure}

\begin{figure}[!ht]
    \centering
    \includegraphics[width=\textwidth]{\detokenize{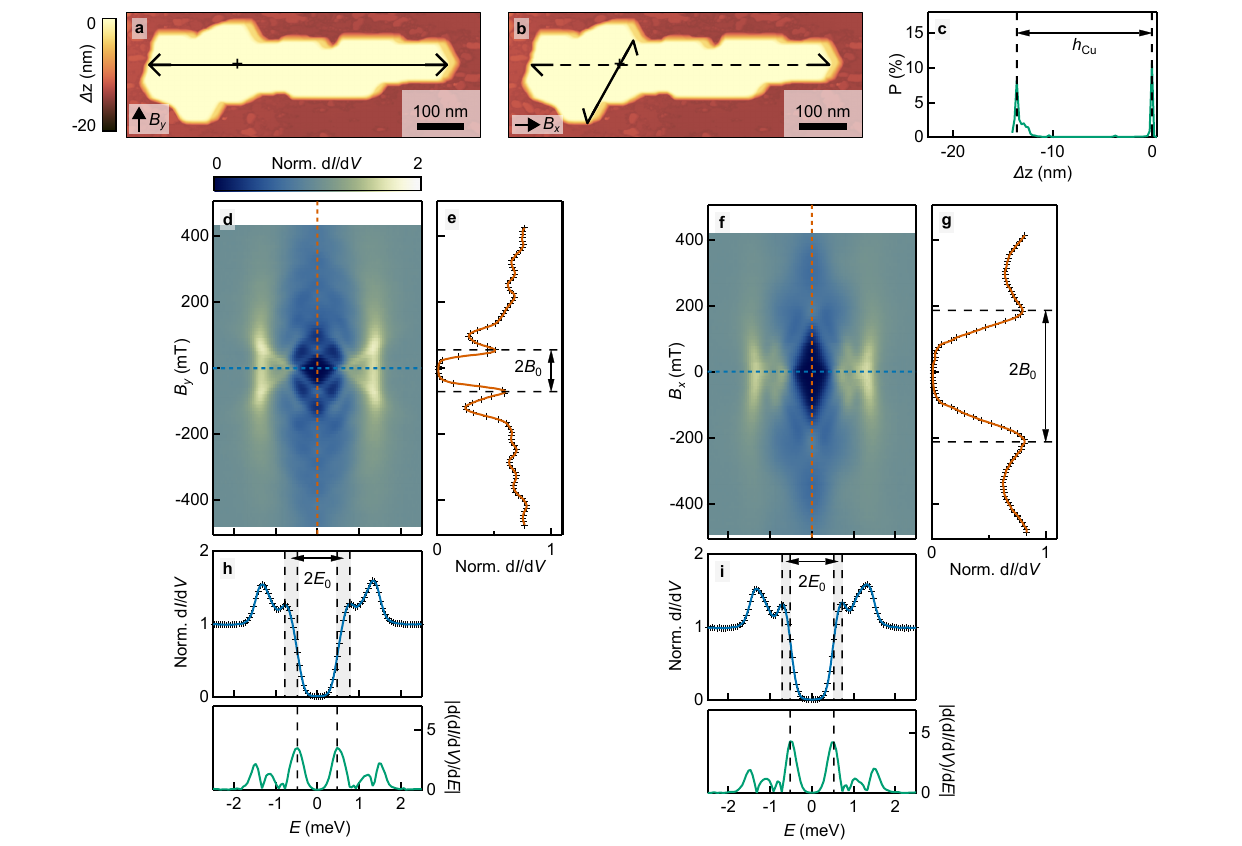}}
    \caption{\textbf{Extraction of \(E_{0}\) and \(B_{0}\) for island D.}
    Island D is shown in Fig.~2c and analysed in Figs.~3 and 4.
    \textbf{a,b} Topographic images for \(B_y\) and \(B_x\). Crosses mark the spectroscopy positions and black lines indicate trajectories associated with the bundle-derived \(\ell\) and \(w\) scales.
    \textbf{c} Height histogram used to determine \(h_{\mathrm{Cu}}\).
    \textbf{d,f} Normalised \(\mathrm{d}I/\mathrm{d}V\) maps for the same field directions. Dashed lines mark the zero-bias and zero-field cuts.
    \textbf{e,g} Zero-bias conductance cuts. Selected maxima define \(B_0\); arrows denote \(2B_0\).
    \textbf{h,i} Zero-field spectra and absolute first derivatives. Dashed vertical lines and shaded brackets define \(E_0^{\mathrm{low}}\) and \(E_0^{\mathrm{high}}\); arrows indicate \(2E_0\).
    Parameters: \textbf{a,b} \(V{=}900\)\,mV, \(I{=}1\)\,nA;
    \textbf{d--i} \(V_\mathrm{stab}{=}500\)\,mV, \(I_\mathrm{stab}{=}500\)\,nA, \(V_\mathrm{mod}{=}50\)\,µV.
    }
    \label{fig:SM_Isl1_E0_B0}
\end{figure}

\begin{figure}[!ht]
    \centering
    \includegraphics[width=\textwidth]{\detokenize{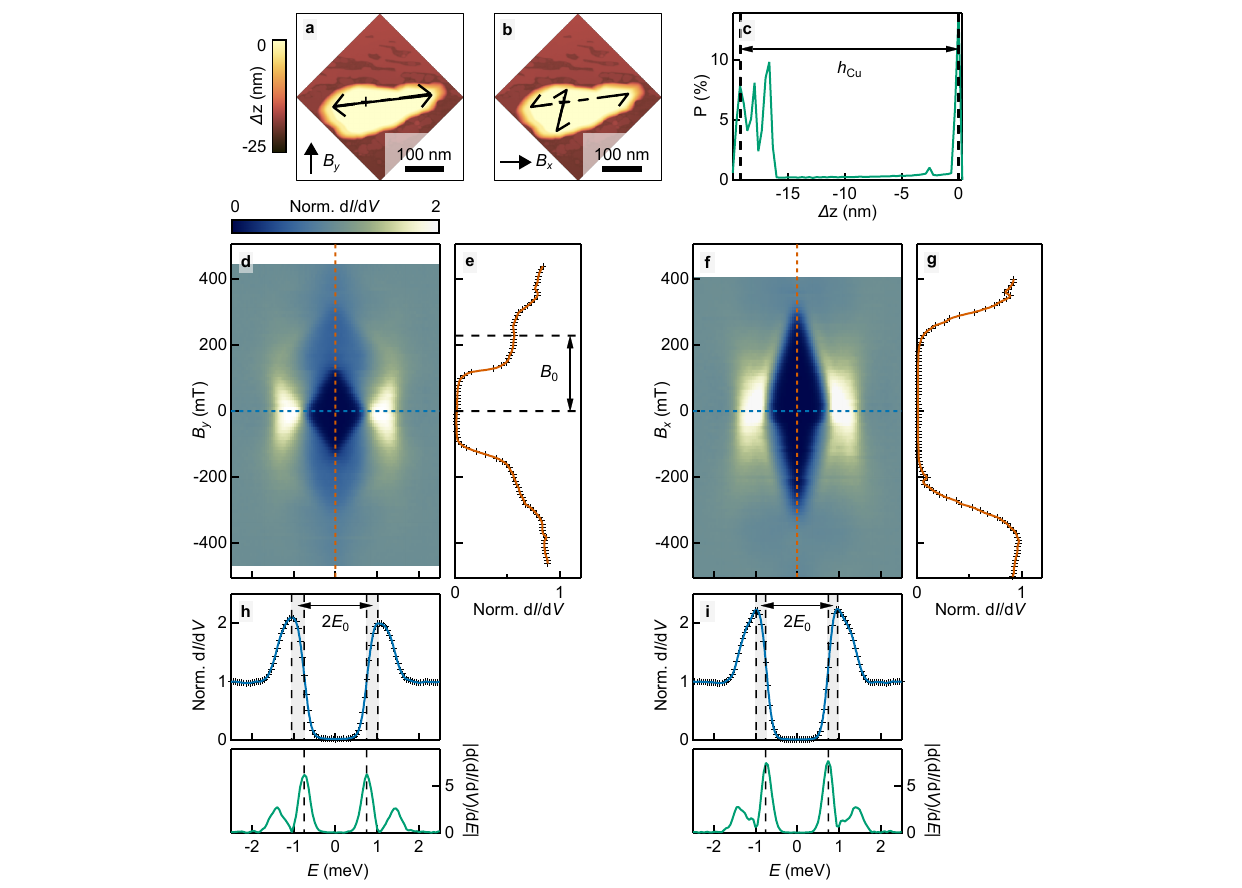}}
    \caption{\textbf{Extraction of \(E_{0}\) and \(B_{0}\) for island E.}
    \textbf{a,b} Topographic images for \(B_y\) and \(B_x\). Crosses mark the spectroscopy positions and black lines indicate trajectories associated with the bundle-derived \(\ell\) and \(w\) scales.
    \textbf{c} Height histogram used to determine \(h_{\mathrm{Cu}}\).
    \textbf{d,f} Normalised \(\mathrm{d}I/\mathrm{d}V\) maps for the same field directions. Dashed lines mark the zero-bias and zero-field cuts.
    \textbf{e,g} Zero-bias conductance cuts. Arrow indicates \(B_0\) based on the selected maximum. The \(B_x\) trace is shown for completeness; jumps in the zero-bias conductance prevent a robust \(B_0\) extraction.
    \textbf{h,i} Zero-field spectra and absolute first derivatives. Dashed vertical lines and shaded brackets define \(E_0^{\mathrm{low}}\) and \(E_0^{\mathrm{high}}\); arrows indicate \(2E_0\).
    Parameters: \textbf{a,b} \(V{=}500\)\,mV, \(I{=}1\)\,nA;
    \textbf{d--i} \(V_\mathrm{stab}{=}500\)\,mV, \(I_\mathrm{stab}{=}500\)\,nA, \(V_\mathrm{mod}{=}50\)\,µV.
    }
    \label{fig:SM_Isl2_E0_B0}
\end{figure}

\section{One-dimensional Usadel benchmark}

The one-dimensional Usadel calculation is used as a diffusive benchmark, not as a model of the ballistic surface-state trajectory ensemble. We choose boundary conditions that favour a robust proximity minigap and compare the resulting field evolution with the experiment and the trajectory-ensemble calculation.

To construct the benchmark, we solve a one-dimensional phase-biased Usadel problem for an effective SNS junction. The retarded Green function is parameterised by
\begin{equation*}
    g(E,x)=\cos\theta(E,x),
    \qquad
    f(E,x)=\sin\theta(E,x)\,\mathrm{e}^{\mathrm{i}\chi(E,x)} .
\end{equation*}
In the normal region we solve
\begin{equation*}
    \frac{\hbar D}{2}
    \left[
        \partial_x^2\theta
        -
        \left(\partial_x\chi\right)^2
        \sin\theta\cos\theta
    \right]
    +
    \mathrm{i}(E+\mathrm{i}\eta)\sin\theta
    =
    0 ,
    \label{eq:usadel_theta}
\end{equation*}
and
\begin{equation*}
    \partial_x
    \left[
        \sin^2\theta\,\partial_x\chi
    \right]
    =
    0 .
    \label{eq:usadel_chi}
\end{equation*}
Here \(D=v_\mathrm{F}\ell_{\mathrm{eff}}/2\), \(\ell_{\mathrm{eff}}\) is the effective elastic mean free path, and \(\eta\) is the intrinsic broadening. We use the same \(\Delta_{\mathrm{eff}}\), intrinsic broadening and experimental-resolution convolution as in the trajectory-ensemble calculation.

The S--N interfaces are taken to be highly transparent, consistent with the small BTK barrier used in the trajectory-ensemble calculation. We therefore impose rigid BCS boundary conditions
\begin{equation*}
    \theta(0)=\theta(L_D)=\theta_S(E),
    \qquad
    \cos\theta_S(E)=
    \frac{E+\mathrm{i}\eta}{\sqrt{(E+\mathrm{i}\eta)^2-\Delta_{\mathrm{eff}}^2}},
\end{equation*}
and
\begin{equation*}
    \chi(0)=-\frac{\phi}{2},
    \qquad
    \chi(L_D)=+\frac{\phi}{2}.
\end{equation*}
The local density of states is
\begin{equation*}
    N(E,x_{\mathrm{STS}})
    =
    N_0\,\mathrm{Re}
    \left[
        \cos\theta(E,x_{\mathrm{STS}})
    \right].
\end{equation*}

The effective diffusion length \(L_D\) is obtained from the ray-traced ensemble of boundary-to-boundary chord lengths \(\ell_n\) through the STS position
\begin{equation*}
    L_D=\sqrt{\langle \ell_n^2\rangle}.
\end{equation*}
The effective probe coordinate \(x_{\mathrm{STS}}\) is obtained from the average fractional position of the tunnelling point along the same chord ensemble.

The magnetic phase lever arm is treated separately. For each trajectory with hit points \(\mathbf r_{1,n}\) and \(\mathbf r_{2,n}\), we define the transverse projection perpendicular to the applied in-plane field
\begin{equation*}
    w_n=
    \left|
    \left(\mathbf r_{2,n}-\mathbf r_{1,n}\right)
    \cdot \hat{\mathbf e}_\perp
    \right| .
\end{equation*}
For the diffusive comparison we use the RMS transverse phase lever arm
\begin{equation*}
    L_{\perp}=\sqrt{\langle w_n^2\rangle}.
\end{equation*}
The phase bias is then
\begin{equation*}
    \phi_{\mathrm{raw}}(B)=
    \frac{2\pi}{\Phi_0}B\,h_{\mathrm{eff}}\,L_{\perp} .
\end{equation*}
Using the ideal SNS symmetry
\(N(E,\phi)=N(E,-\phi)=N(E,2\pi-\phi)\), the calculation is performed with the folded phase
\begin{equation*}
    \phi =
    \left|
    \mathrm{Arg}
    \left[
        \mathrm{e}^{\mathrm{i}\phi_{\mathrm{raw}}}
    \right]
    \right|,
    \qquad
    0\leq\phi\leq\pi .
\end{equation*}

The mean free path was estimated from impurity and boundary scattering using Matthiessen's rule
\begin{equation*}
    \frac{1}{\ell_{\mathrm{eff}}}
    =
    \frac{1}{\ell_{\mathrm{imp}}}
    +
    \frac{1}{\ell_{\mathrm{edge}}}.
\end{equation*}
From approximately 30 visible scatterers in a
\(100~\mathrm{nm}\times100~\mathrm{nm}\) STM topograph, we estimate
\begin{equation*}
    n_{\mathrm{imp}}\simeq3.0\times10^{-3}~\mathrm{nm}^{-2}.
\end{equation*}
For unitary scatterers in a two-dimensional metal,
\begin{equation*}
    \sigma_{\mathrm{tr}}^{2D}\simeq \frac{4}{k_\mathrm{F}}
    =
    \frac{2}{\pi}\lambda_\mathrm{F} .
\end{equation*}
Using \(\lambda_\mathrm{F}\simeq2.83~\mathrm{nm}\), we obtain
\begin{equation*}
    \ell_{\mathrm{imp}}\simeq185~\mathrm{nm}.
\end{equation*}
The finite island boundary is included as an additional diffuse scattering channel. For the island used in the benchmark, the STM-mask area and perimeter give
\begin{equation*}
    \ell_{\mathrm{edge}}\simeq77.6~\mathrm{nm}.
\end{equation*}
Together with the impurity mean free path this gives
\begin{equation*}
    \ell_{\mathrm{eff}}\simeq54.7~\mathrm{nm},
    \qquad
    D=\frac{1}{2}v_\mathrm{F}\ell_{\mathrm{eff}}.
\end{equation*}

Within this benchmark, \(L_D\) controls the Thouless/minigap scale and zero-field line shape, whereas \(L_{\perp}\) maps the applied in-plane field onto an effective phase bias through the scale parameter \(h_{\mathrm{eff}}\). Since \(h_{\mathrm{eff}}\) is not independently fixed, the field scale can be adjusted within a plausible range. The more relevant comparison is therefore the spectral evolution and line shape rather than the absolute collapse field alone.

\begin{figure}[!ht]
    \centering
    \includegraphics[width=0.9\textwidth]{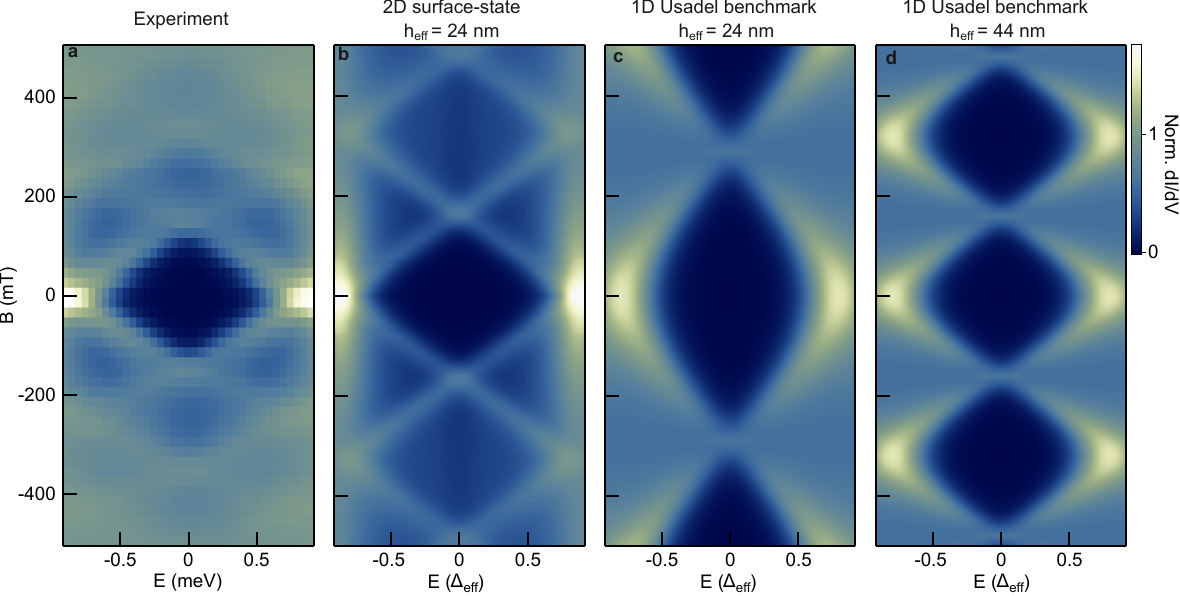}
    \caption{
    \textbf{Comparison with a one-dimensional diffusive Usadel benchmark.}
    \textbf{a} Experimental spectrum reproduced from Fig.~1d.
    \textbf{b} Two-dimensional surface-state trajectory simulation.
    \textbf{c} One-dimensional Usadel benchmark with \(h_{\mathrm{eff}}=24~\mathrm{nm}\), as used in \textbf{b}.
    \textbf{d} One-dimensional Usadel benchmark with \(h_{\mathrm{eff}}=44~\mathrm{nm}\), chosen to match the first low-field zero-bias collapse. The Usadel benchmarks show smooth minigap suppression but do not reproduce the structured spectral evolution seen in the experiment and trajectory simulation. All panels use a common colour scale.
    Parameters: \textbf{a} \(V_\mathrm{stab}{=}500\)\,mV, \(I_\mathrm{stab}{=}500\)\,nA, \(V_\mathrm{mod}{=}50\)\,µV.
    }
    \label{fig:SI_ExpVSSimulationVSUsadel}
\end{figure}

\section{Surface-state and bulk-derived tunnelling channels}

To estimate the possible contribution of bulk-derived Cu states, we applied the same trajectory-based phase analysis to a three-dimensional bulk-derived channel and compared it with the two-dimensional surface-state channel. The two channels were assigned different band and gap parameters, as described in Section~\ref{sec:SI_Model}.

The relative channel weights were estimated from large-bias conductance. After division by a slowly varying linear tunnel background, the averaged spectrum was described by a minimal two-channel decomposition: a bulk-derived contribution and a step-like surface-state contribution at \(E_{\mathrm{SS}}\). This gives comparable tunnelling weights near the Fermi level, approximately \(60\%\) surface-state and \(40\%\) bulk-derived contribution.

For the bulk-derived channel, the relevant three-dimensional geometry cannot be read directly from the STM topography because the apparent side-wall profile may be broadened by the tip shape. We therefore used the STM-imaged top-surface outline, island height and Cu(111) lattice geometry to estimate a trapezoidal island cross-section.

The dominant short-length weight of the bulk-derived ensemble arises from near-normal trajectories at the Cu/Nb interface. These trajectories mainly traverse the Cu thickness and return, producing lengths of order \(2h\). For the \(15~\mathrm{nm}\)-high island considered here, this gives \(\ell_n\sim2h\approx30~\mathrm{nm}\). Their lateral displacement is small, and hence their transverse projection \(w_n\) is also small.

This short length scale places the dominant bulk-derived contribution close to the Nb gap edge. For the bulk-derived parameters used here, \(\hbar v_{\mathrm{F},\mathrm{3D}}/\Delta_{\mathrm{3D}}\simeq795~\mathrm{nm}\), so the dominant near-normal paths have \(\ell_n/(\hbar v_{\mathrm{F},\mathrm{3D}}/\Delta_{\mathrm{3D}})\approx0.04\). Their zero-field Andreev levels therefore remain close to the gap edge, \(E\simeq\Delta_{\mathrm{3D}}\simeq1.3~\mathrm{meV}\).

Longer bulk-derived trajectories require more grazing incidence and larger lateral displacement. In the trapezoidal geometry, such trajectories are often redirected toward the Nb interface by specular reflection at the side wall, which limits the long-\(\ell_n\) and large-\(w_n\) tails. The histograms in Supplementary Fig.~\ref{Sfig:3DbulkRayTracing} therefore reflect two effects: dominant near-normal trajectories produce the main short-\(\ell_n\), small-\(w_n\) weight, while side-wall reflections suppress extended multi-bounce bulk paths.

The weighted distributions in Supplementary Fig.~\ref{Sfig:3DbulkRayTracing} are the geometric input to the bulk-derived spectral calculation. They show that, compared with the two-dimensional surface-state ensemble, the bulk-derived ensemble has little weight at large \(w_n\) or large \(\ell_n\). Since only geometric weights are included, this likely overestimates the contribution of strongly oblique bulk-like trajectories, which should be further suppressed in STM tunnelling.

The resulting spectral comparison in Supplementary Fig.~\ref{Sfig:3DbulkLDOS} shows that the bulk-derived channel can add appreciable conductance, especially near the Nb gap edge and at higher energies. However, because it has the larger gap scale \(\Delta_{\mathrm{3D}}\), a different effective in-plane-field lever arm and dominant short, near-normal trajectories, it does not reproduce the low-field zero-bias collapse. That response is captured by the two-dimensional Cu(111) surface-state trajectory ensemble. Thus, including the bulk-derived contribution changes the high-energy and gap-edge spectral weight but does not alter the assignment of the low-field spectral organisation to confined two-dimensional surface-state trajectories.

\begin{figure}[ht!]
    \centering
    \includegraphics[width=0.8\textwidth]{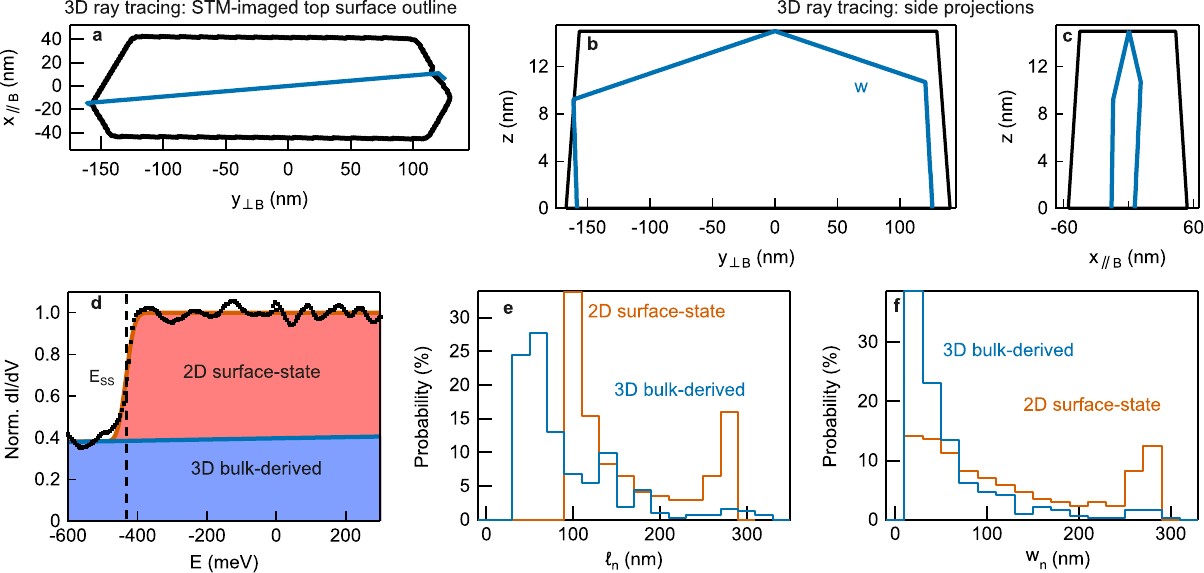}
    \caption{
    \textbf{Tunnelling weights and trajectory scales of two-dimensional surface-state and bulk-derived Cu channels.}
    \textbf{a--c} STM-imaged top-surface outline and side projections used for the bulk-derived ray tracing. The blue line marks the transverse projection associated with the bundle-derived \(w\) scale; the black outline marks the island boundary in the corresponding projection.
    \textbf{d} Large-bias conductance averaged over the island after division by a linear tunnel background. The corrected spectrum is compared with a minimal two-channel decomposition into a bulk-derived contribution and a step-like surface-state contribution at \(E_{\mathrm{SS}}\).
    \textbf{e,f} Weighted distributions of trajectory length \(\ell_n\) and transverse projection \(w_n\) for the ray-traced surface-state and bulk-derived channels. The bulk-derived ensemble is dominated by short, near-normal trajectories with small \(w_n\), whereas the surface-state ensemble contains pronounced large-\(w_n\) and large-\(\ell_n\) tails.
    }
    \label{Sfig:3DbulkRayTracing}
\end{figure}

\begin{figure}[ht!]
    \centering
    \includegraphics[width=0.8\textwidth]{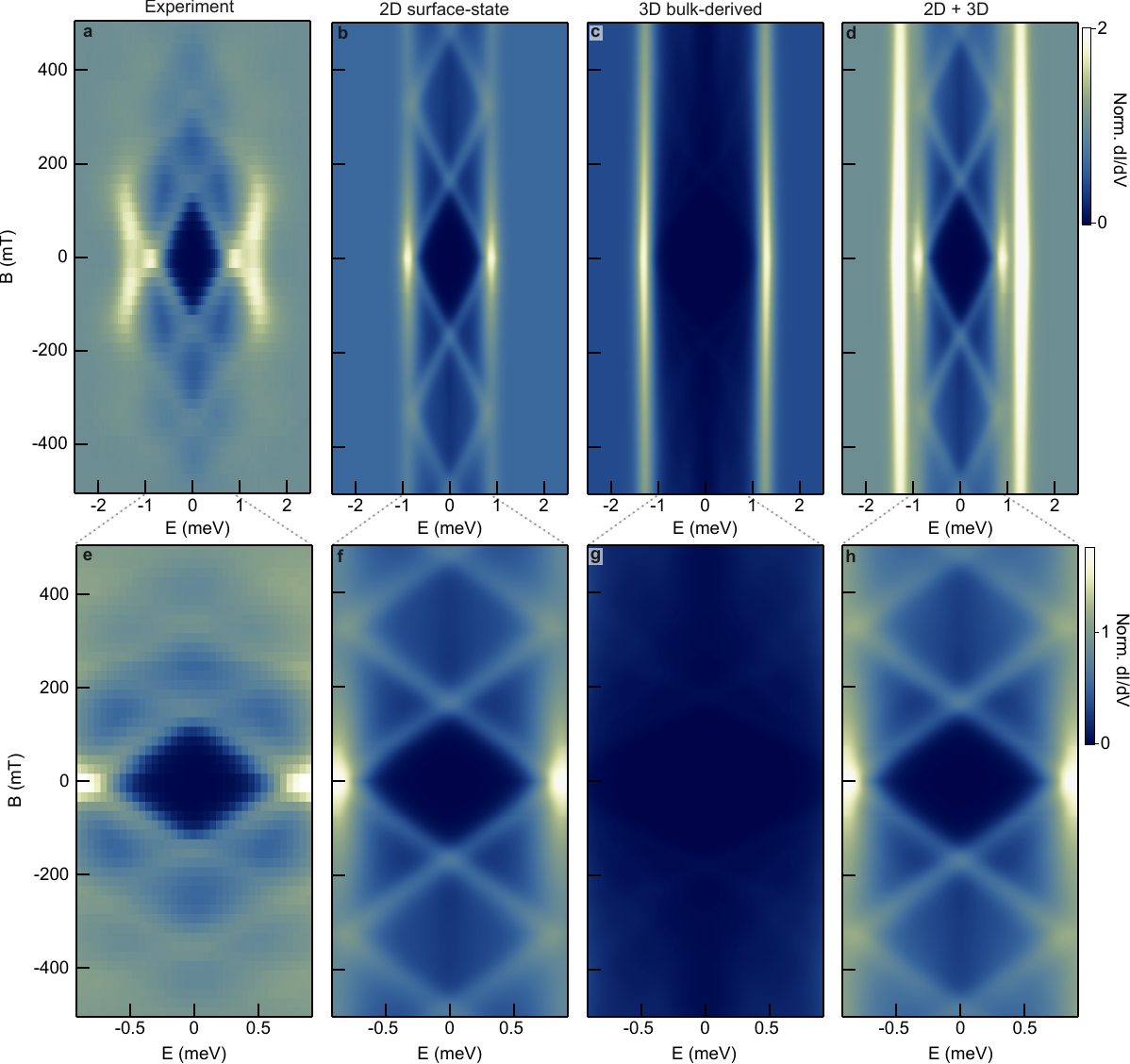}
    \caption{
    \textbf{Spectral contributions of surface-state and bulk-derived channels.}
    Experimental data in \textbf{a,e} are reproduced from Fig.~1d and compared with trajectory-ensemble simulations for the surface-state channel, bulk-derived channel and weighted sum.
    \textbf{a--d} Wide energy-window comparison. The bulk-derived contribution appears mainly near the Nb gap edge, \(\Delta_{\mathrm{3D}}\simeq1.3~\mathrm{meV}\), whereas the surface-state channel captures the central low-field spectral collapse.
    \textbf{e--h} Corresponding zoom into the low-energy window bounded by \(\pm\Delta_{\mathrm{eff}}\). The spectral organisation in this range is dominated by the surface-state trajectory ensemble. The bulk-derived channel contributes weakly because most of its spectral weight lies outside this window.
    Within each row, all panels are shown on a common colour scale. The strongest intensities in \textbf{d} exceed the displayed range and are clipped to preserve the visibility of low-energy features.
    Parameters: \textbf{a} \(V_\mathrm{stab}{=}500\)\,mV, \(I_\mathrm{stab}{=}500\)\,nA, \(V_\mathrm{mod}{=}50\)\,µV.
    }
    \label{Sfig:3DbulkLDOS}
\end{figure}

\section{Interface-layer control spectra and experimental energy resolution}
\label{sec:SM_interface_layer_control}

We performed control spectroscopy on the interface layer surrounding the Cu islands, where the confined Cu(111) surface state is absent. Field-dependent spectra were acquired for two in-plane field directions, \(B_y\) and \(B_x\), using the same bias range as for the island measurements.

The resulting \((\mathrm{d}I/\mathrm{d}V)_{\mathrm{norm}}\) maps show the superconducting gap and the field evolution of the coherence peaks, but not the low-energy dispersing Andreev features or low-field zero-bias collapse observed on the Cu islands. This supports assigning the field-dependent low-energy spectra in the main text to the proximitised Cu(111) surface state of the islands.

A zero-field spectrum on the same interface layer was used to determine the experimental energy resolution. Fitting the superconducting gap with a thermally broadened Dynes density of states gives an effective electronic temperature \(T_{\mathrm{eff}}=0.86~\mathrm{K}\). This value was used for the experimental-resolution convolution in the trajectory-ensemble and benchmark calculations.

\begin{figure}[!ht]
    \centering
    \includegraphics[width=\textwidth]{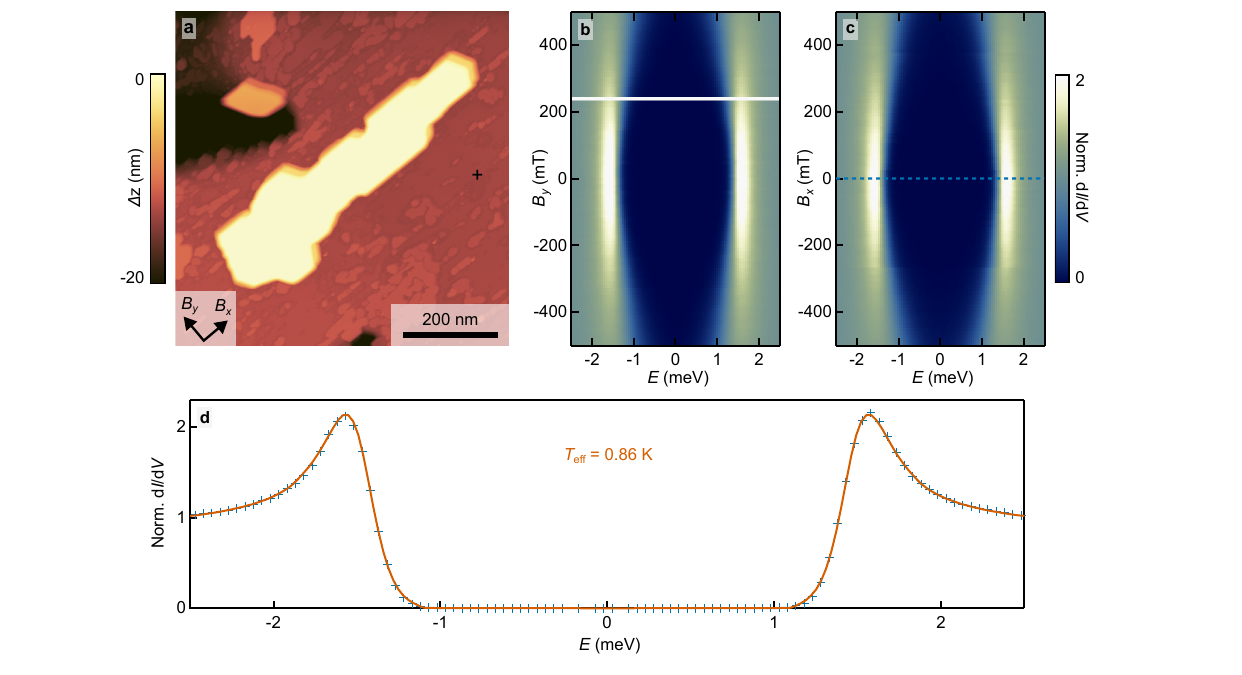}
    \caption{\textbf{Control measurement on the interface layer.}
    \textbf{a} Topographic image of the island D and the surrounding interface layer. The cross marks the position at which the field-dependent spectra were acquired.
    \textbf{b,c} Normalised \(\mathrm{d}I/\mathrm{d}V\) maps measured on the interface layer for \textbf{b} \(B_y\) and \textbf{c} \(B_x\). The spectra show the superconducting gap and field evolution of the coherence peaks, but not the low-energy dispersing Andreev features observed on the Cu islands.
    \textbf{d} Spectrum of the interface layer at zero field, marked by the dashed blue line in \textbf{c}, together with a thermally broadened Dynes fit. The fit gives \(T_{\mathrm{eff}}=0.86~\mathrm{K}\), which was used for the experimental-resolution convolution throughout the simulations.
    Parameters: \textbf{a} \(V{=}-900\)\,mV, \(I{=}-0.1\)\,nA;
    \textbf{b,c} \(V_\mathrm{stab}{=}500\)\,mV, \(I_\mathrm{stab}{=}500\)\,nA, \(V_\mathrm{mod}{=}50\)\,µV.
    }
    \label{fig:SM_interface_layer_control}
\end{figure}